\documentclass[a4paper,12pt]{article}
\pdfoutput=1 % if your are submitting a pdflatex (i.e. if you have
\usepackage{jheppub} % for details on the use of the package, please
\usepackage{tabularx,subcaption,fancyhdr,mathtools,physics}
\usepackage[makeroom]{cancel}
\usepackage{bm}
\usepackage{csquotes}
\usepackage[nottoc,notlot,notlof]{tocbibind}
\usepackage{float}
\usepackage{relsize}
\usepackage{hyperref}
\usepackage{color,latexsym, array,multirow,url,verbatim, enumerate,cancel}
\DeclareUnicodeCharacter{2212}{-}
\hypersetup{colorlinks=true,linkcolor=blue,citecolor=magenta,filecolor=magenta,
	urlcolor=cyan}
\def\tcb{\textcolor{blue}}

\def \lspone{\widetilde\chi_1^0}
\def \mlspone{m_{\lspone}}

\newcommand{\beq}{\begin{equation}}
\newcommand{\eeq}{\end{equation}}
\def\bea{\begin{eqnarray}}
\def\eea{\end{eqnarray}}

\title{\boldmath Bilinear R-parity violating supersymmetry under the light of neutrino oscillation, higgs and flavor data}

\author[a]{Arghya Choudhury,} 
\author[b]{Sourav Mitra,}
\author[a]{Arpita Mondal,}
\author[c]{and Subhadeep Mondal}

\affiliation[a]{Department of Physics, Indian Institute of Technology Patna, Bihar - 801106, India}
\affiliation[b]{Surendranath College, 24/2 M. G. Road, Kolkata, West Bengal - 700009, India}
\affiliation[c]{Department of Physics, SEAS Bennett University, Greater Noida, Uttar Pradesh  -201310, India}
\emailAdd{arghya@iitp.ac.in}
\emailAdd{hisourav@gmail.com}
\emailAdd{arpita\_1921ph15@iitp.ac.in (corresponding author)}
\emailAdd{subhadeep.mondal@bennett.edu.in}

\abstract{In this work, we explore a well motivated beyond the Standard Model scenario, namely, R-parity violating Supersymmetry, in the context of light neutrino masses and mixing. We assume that the R-parity is only broken by the lepton number violating bilinear term. We try to fit two non-zero neutrino mass square differences and three mixing angle values obtained from the global $\chi^2$ analysis of neutrino oscillation data. We have also taken into account the updated data of the standard model (SM) Higgs mass and its coupling strengths with other SM particles from LHC Run-II along with low energy flavor violating constraints like rare b-hadron decays. We have used a Markov Chain Monte Carlo (MCMC) analysis to constrain the new physics parameter space. While doing so, we ensure that all the existing collider constraints are duly taken into account. Through our analysis, we have derived the most stringent constraints possible to date with existing data on the 9 bilinear R-parity violating parameters along with $\mu$ and $\tan\beta$. 
We further explore the possibility of explaining the anomalous muon~(g~-~2) measurement staying within the parameter space allowed by neutrino, Higgs and flavor data while satisfying the collider constraints as well. We find that there still remains a small sub-TeV parameter space where the required excess can be obtained.}

\begin{document} 
\maketitle

\flushbottom

\setlength{\parskip}{1.0em}

\section{Introduction}
\label{sec:intro}

Neutrino oscillation is one of the most robust indications towards the existence of physics beyond the standard model (BSM). Over the years, multiple experiments have been studying the neutrino oscillation phenomena, see e.g., \cite{Borexino:2013zhu,KamLAND:2013rgu,RENO:2018dro,DayaBay:2018yms,Super-Kamiokande:2019gzr,T2K:2018rhz,NOvA:2019cyt}. Their measurement of two mass square differences and three mixing angles imply significant mixing among the three light neutrino states of which at least two must have non-zero masses \cite{deSalas:2020pgw}. The standard model (SM) \cite{GLASHOW1961579,PhysRevLett.19.1264,Salam:1968rm,GELLMANN1964214} or 
the R-parity\footnote{R-parity is defined as $R_p=(-1)^{(3B-L+2S)}$, where B, L and S are baryon number, lepton number and spin quantum number of the particle respectively.} conserving minimal supersymmetric standard model (MSSM)~\cite{Drees:2004jm, Baer:2006rs, Martin:1997ns} cannot address the neutrino oscillation phenomena. The light neutrino masses and mixing can be generated by simple seesaw extensions of the SM, which are different manifestations of the dimension-5 Weinberg operator \cite{Weinberg:1979sa, Weinberg:1980bf}. In Type-I seesaw, we add right-handed singlet fermions in the model, in Type-II seesaw, we add fermionic triplets and in Type-III seesaw the objective is achieved by adding scalar triplets to SM \cite{Minkowski:1977sc,Gell-Mann:1979vob,Schechter:1980gr,Mohapatra:1979ia,Schechter:1981cv}. In R-parity violating (RPV) MSSM \cite{Dreiner:1997uz, Barbier:2004ez, Banks:1995by,Grossman:1998py,Davidson:2000uc,Davidson:2000ne,Borzumati:1996hd,Mukhopadhyaya:1998xj,PhysRevD.59.091701,PhysRevD.61.055006,Allanach:2007qc,Allanach:2011de,Grossman:1997is,Dreiner:1991pe,Dercks:2017lfq,Bose:2014vea,Datta:2009dc,Das:2005mr,Mitsou:2015kpa,Cohen:2019cge} scenario, one can explain neutrino oscillation phenomena without incorporating the Weinberg operator. In the light of the updated neutrino oscillation data and other relevant constraints, it is worth revisiting the scenario to gauge their impact on the RPV couplings. 

R-parity conserving MSSM is more widely studied in literature because it offers a natural dark matter candidate in the form of the lightest supersymmetric particle (LSP) which cannot decay further and is therefore stable\footnote{It is also possible to have a dark matter candidate within RPV scenario in the form of a very long-lived neutralino, gravitino or axino. See e.g., \cite{Barbier:2004ez,Colucci:2018yaq,Bae:2017tqn}}. However, one has to incorporate R-parity conservation by hand to achieve that. The symmetry principles to write the Lagrangian allow us to add four RPV terms in the superpotential as below: 
\begin{equation}
\label{eq:rpv_potential}
W_{\cancel{R}_p} = \epsilon_i \hat{L}_i \hat{H}_u + \frac{1}{2}\lambda_{ijk}\hat{L}_i\hat{L}_j\hat{E}_k^c + \frac{1}{2}\lambda^{\prime}_{ijk}\hat{L}_i\hat{Q}_j\hat{D}_k^c + \frac{1}{2}\lambda^{\prime\prime}_{ijk}\hat{U}_i^c\hat{U}_j^c\hat{D}_k^c 
\end{equation}

The bilinear term,  $\epsilon_i \hat{L}_i \hat{H}_u$,  and the next two trilinear terms containing $ \lambda$, $\lambda^{\prime}$  in the Eq.~\ref{eq:rpv_potential} each violates lepton number by one unit and the last $\lambda^{\prime\prime}$ term violates the baryon number by one unit. Here $\hat{L}_i$ ($\hat{E}_k$) corresponds to the left (right) handed lepton supermultiplet and $\hat{H}_u$ is the up-type Higgs supermultiplet. 
$\hat{Q}_j$, $\hat{U}_j$ ($\hat{D}_k$)  represent left-handed doublet and right-handed singlet up-type 
(down-type) quark supermultiplet respectively. 
One can generate non-zero light neutrino masses through the trilinear $\lambda$ or $\lambda^{\prime}$ couplings at one-loop~\cite{Rakshit:2004rj,Grossman:2003gq}. Note that, with these couplings, the neutrinos are still massless at tree level. The bilinear term is capable of generating one neutrino mass at tree level \cite{Grossman:1997is,Rakshit:2004rj,Grossman:2003gq}. However, one also needs to take into account the one-loop contributions to explain the oscillation data. The trilinear couplings need not be non-zero in that case from the perspective of light neutrino mass generation. Note that, the bilinear RPV (bRPV) terms can exist even in the absence of the trilinear terms in the theory and the trilinear couplings can be generated starting from the bilinear couplings \cite{Roy:1996bua}. On the other hand, if one starts from only trilinear RPV terms, the bilinear RPV couplings can be generated through renormalisation group evolution at a different energy scale \cite{deCarlos:1996ecd,Nardi:1996iy}. Understandably, neutrino masses and mixing angles lead to constraints on the trilinear couplings \cite{Allanach:1999ic}. Bilinear RPV can be assumed to be the fundamental theory and hence, in this work, we only focus on non-zero values of these couplings keeping all trilinear RPV couplings zero. 
Note that, in the alignment of bilinear coupling parameters ($\epsilon_i$) and sneutrino vev ($v_i$), which arises naturally in the framework of horizontal symmetries, the three light mass eigenstates of $L_i$ correspond to the three light neutrinos ~\cite{Banks:1995by,Allanach:2003eb}. To achieve this, the alignment between soft coupling parameters, $B_i$ and $\epsilon_i$ is also required i.e., $B_i \propto \epsilon_i$~\cite{Banks:1995by,Allanach:2003eb}. In such cases, both the bilinear term and the soft breaking bilinear term can be rotated away by the field redefinition of $L_i$ and $H_u$.

Although the contribution of bilinear RPV couplings towards neutrino masses and mixings has been studied in the past \cite{Hempfling:1995wj,Hirsch:2000ef,Hundi:2011si,Diaz:2014jta}, a detailed statistical analysis that can highlight the allowed parameter space is missing in the existing literature. There has been some effort to constrain the bilinear RPV couplings from neutrino physics perspective \cite{Hempfling:1995wj,Hirsch:2000ef,Hirsch:2000jt,Abada:2001zh,Diaz:2004fu,Hundi:2011si,deCampos:2012pf,Diaz:2014jta,Gozdz:2008zz} but they are either not very generalized or simply inadequate in the light of new oscillation data \cite{deSalas:2020pgw}. Moreover, adding RPV couplings leads to a wide range of phenomenological implications \cite{Barbier:2004ez}. One needs to take into account the modified bounds on the SUSY particles which can now decay exclusively into SM particles. Because of the presence of the bilinear RPV term, the neutral and charged Higgs states now can mix with the sneutrinos and charged sleptons respectively. Therefore, one needs to carefully check the SM Higgs coupling strengths in the light of the updated dataset \cite{cms_web1,ATLAS:2021vrm} which can further put constraints on the RPV parameters. In this study, we take into account all these possibilities.

New measurement of muon magnetic moment at Fermilab has slightly changed the existing world average, which shows a $4.2\sigma$ deviation\footnote{The recent QCD lattice simulation of the Hadronic Vacuum Polarization (HVP) term by the BMW collaboration~\cite{Borsanyi:2020mff}, CLS/Mainz group~\cite{Ce:2022kxy}, 
Extended Twisted Mass Collaboration (ETMC)~\cite{ExtendedTwistedMass:2022jpw} and the preliminary results of the CDM-3 detector~\cite{CMD-3:2023alj}  indicate that the discrepancy between the observed and predicted values of muon (g~-~2) will be smaller and less significant.} at present from SM 
prediction~\cite{Muong-2:2006rrc,Muong-2:2021ojo}. 
\begin{equation}
\label{eq:muon_g}
\Delta a_{\mu} = a_{\mu}^{Exp} - a_{\mu}^{SM} = (25.1 \pm 5.9) \times10^{-10} 
\end{equation}
Sneutrino-chargino and slepton-neutralino loops have been studied extensively in this context and it is evident that given the present collider constraints, it is quite difficult to achieve the required excess within the framework of the MSSM. 
One region of sub-TeV allowed parameter space still relevant in this context 
is the compressed region where LSP-NLSP mass difference is very small~\cite{Endo:2021zal}.   
%One region of sub-TeV allowed parameter space still relevant in this context is the compressed region where at least one of the \tcb{sneutrino-chargino} and slepton-neutralino pair is lying at the bottom of the SUSY spectrum and their mass gap is less than~$\sim 50$~GeV \cite{}. 
These kinds of compressed regions are difficult to probe owing to the poor detection prospect of the final state particles. 
In the RPV context, however, these particles can decay further into SM particles if the RPV couplings are large enough for a prompt decay. That leads to new collider constraints obtained from direct searches at the LHC \cite{ATLAS:2021yyr,Barman:2020azo,ATLAS:2021moa,ATLAS:2023lfr,ATLAS:2014kpx,ATLAS:2014eel,ATLAS:2015rul,ATLAS:2015gky,atlas_web4}. Having said that, there are some additional contributions to the muon~(g~-~2) in the RPV framework owing to the mixing between charged higgs- sleptons, neutral higgs- sneutrinos, charginos- charged leptons and neutralino- neutrinos. 
In the present context, it is worth a look since RPV couplings and some relevant particle masses, e.g., that of sneutrinos and neutralinos have a big role to play in both muon~(g~-~2) calculation and generation of light neutrino masses and mixing. We, therefore, explore the possibility of explaining the muon~(g~-~2) excess over the SM contribution within this framework while simultaneously satisfying all other experimental constraints.

The presence of the bilinear RPV term results in a lepton number violation by one unit. A sneutrino state therefore can now acquire non-zero vacuum expectation value (VEV). All three sneutrino VEVs along with the three $\epsilon_i$ parameter and their corresponding soft terms are crucial in fitting the neutrino oscillation data. Moreover, because of the sneutrino and neutral Higgs mixing, the Higgs sector parameters like trilinear stop coupling ($A_t$), the ratio of up and down type Higgs vacuum expectation values (${\rm tan}\beta$) and the $\mu$ parameter are also relevant to our objective. This results in a substantially bigger set of unknown input parameters and as a result, a conventional random scan does not produce the coveted results. We, therefore, adopt a Markov Chain Monte Carlo (MCMC) algorithm to sample the parameter space. The main objective of this work is to take advantage of the current observational data in order to constrain the model parameters by a robust statistical analysis. The method described here is not only able to enhance our understanding of the importance of currently available data on the modeling of R-parity violating supersymmetry (RPV SUSY), but also explores the ability to use this technique for upcoming experiments. We concentrate mostly on the neutrino and Higgs sector observables apart from some relevant flavor observables to locate the favored parameter region. While doing so, we ensure that the collider constraints on supersymmetric parameters are also taken into account. We have kept the parameter space as generalized as possible albeit with some simplified assumptions on the soft masses of charged sleptons, sneutrinos and squarks.
   
The plan/structure of this paper is as follows. In Sec.\ref{sec:model} we briefly discuss the generation of light neutrino masses and mixings in the context of bRPV SUSY model. In Sec.\ref{sec:num} we first mention the constraints coming from the global analysis of neutrino oscillation data, then Higgs mass and the constraints from Higgs signal strength measurements. In addition, we have considered the flavor constraints arising from rare B decays. We also briefly discuss the prescription of the MCMC technique adopted by us to identify the favored parameter region. Sec.\ref{sec:result} contains the results for normal and inverted hierarchy scenarios. We briefly discuss the prospects of addressing the anomalous muon~(g~-~2) issue in Sec.\ref{sec:muon_g} and finally, we present our conclusion in Sec.\ref{sec:conclusion}.
  
%%%%%%%%%%%%%%%%%%%%%%%%%%%%%%
\section{Neutrino Mass and mixing from Bilinear RPV SUSY}
\label{sec:model}
%%%%%%%%%%%%%%%%%%%%%%%%%%%%%%

In the bRPV SUSY model the additional RPV term in the superpotential generates the mixing between the neutrinos and neutralinos \cite{Grossman:2003gq, Barbier:2004ez, Dreiner:1997uz} and is written as:   %add ref of 0311310
\begin{equation}
W_{bRPV} = \epsilon_i \hat{L}_i \hat{H}_u
\label{eq:superpotential}
\end{equation}
Here $\epsilon_i$ ($i= 1, 2, 3$) represents the bRPV mass parameters. The Lagrangian corresponding to the superpotential gives rise to mixing between up-type Higgsino ($\tilde{H}_u^0$) and three light neutrinos ($\nu_{iL}$). It also generates mixing between $\tilde{H}_u^+$ and left-handed leptons ($l_{iL}$). The Lagrangian of the superpotential and the soft term are given by 
\begin{equation}
\begin{split}
\mathcal{L} = \epsilon_i(\tilde{H}_u^0 \nu_{iL} - \tilde{H}_u^+ l_{iL}) + h.c~~ ; ~~~
\mathcal{L}_{soft} = B_i {\tilde{L}}_i  H_u + {\rm h.c}
\end{split}
\end{equation}
where $B_i$ corresponds to the soft bRPV coupling parameter representing the coupling between sneutrino and the neutral Higgs bosons, and ${\tilde{L}}_i$ is the left-handed slepton multiplet.
So, at the tree level, the resultant $7\times7$ neutralino-neutrino mass matrix in the basis of $\psi^0 =\left( \begin{matrix}
\tilde{B} & \tilde{W}_3 &\tilde{H}_d^0 &\tilde{H}_u^0 &\nu_e &\nu_{\mu} &\nu_{\tau} 
\end{matrix}\right )$ 
looks like \cite{Banks:1995by, Nardi:1996iy, Grossman:1998py, Rakshit:2004rj} as 
\begin{equation}
\label{eq:matrix1}
\begin{pmatrix}
M_1 & 0 & -\frac{1}{2}g^{\prime} v_d & -\frac{1}{2} g^{\prime}v_u & -\frac{1}{2}g^{\prime} v_1 & -\frac{1}{2}g^{\prime}v_2 & -\frac{1}{2}g^{\prime}v_3 \\

0 & M_2 & \frac{1}{2}gv_d & -\frac{1}{2}gv_u & \frac{1}{2}gv_1 & \frac{1}{2}gv_2 & \frac{1}{2}gv_3 \\

-\frac{1}{2}g^{\prime}v_d & \frac{1}{2}gv_d & 0 & -\mu & 0 & 0 & 0 \\

\frac{1}{2}g^{\prime}v_u & -\frac{1}{2}gv_u & -\mu & 0 & \epsilon_1 & \epsilon_2 & \epsilon_3 \\

-\frac{1}{2}g^{\prime}v_1 & \frac{1}{2}gv_1 & 0 & \epsilon_1 & 0 & 0 & 0 \\

-\frac{1}{2}g^{\prime}v_2 & \frac{1}{2}gv_2 & 0 & \epsilon_2 & 0 & 0 & 0 \\

-\frac{1}{2}g^{\prime}v_3 & \frac{1}{2}gv_3 & 0 & \epsilon_3 & 0 & 0 & 0 
\end{pmatrix}
\end{equation}
where, $\tilde{B}$ ($\tilde{W}_3$) denotes bino (wino) and $M_1$ ($M_2$) is bino (wino) mass parameter. $v_u$, $v_d$ are the VEVs for up-type and down-type Higgs respectively. 
The sneutrino VEVs are represented by $\langle\tilde{\nu}_i\rangle \equiv v_i (i = 1,2,3)$. 
Diagonalising the above mass matrix gives rise to one non-zero light neutrino mass apart from four massive neutralinos. However, at least one other light neutrino must be massive in order to satisfy the neutrino oscillation data. This is achieved at one-loop level. 

%%%%%%%%%%%%%%%%%%%%%%%%%%%%%%%%%%%%%%%%%%%%%%
\begin{figure}[!htb]
\begin{center}
    \includegraphics[width=0.6\textwidth]{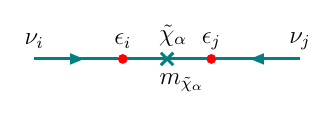}
    \caption{Depiction of generation of light neutrino masses at the tree level.  The cross represents mass insertion.  }
    %Tree level diagram for neutrino mass involving $\epsilon_i$ and mass insertion for the neutralino propagator (represented by the cross) 
    \label{fig:tree}
    \end{center}
\end{figure}
%%%%%%%%%%%%%%%%%%%%%%%%%%%%%%%%%%%%%%%%%%%%%%%

The tree level mass of one of the neutrinos also receives some loop correction and understandably this one happens to be the heaviest of the three neutrinos. The admixture of the three massive neutrinos in the respective mass eigenstates of course depends on the choice of hierarchy. Tree level contribution, $ [m_{\nu}]_{ij}^{\epsilon\epsilon}$, involves ${\epsilon_{i}}$ term, which indicates the mixing between the up-type Higgsino and the neutrino (see Fig.~\ref{fig:tree}). There are two different loop contributions to neutrino masses from this model, namely, the BB loop, and the $\epsilon$B loop \cite{Davidson:2000uc, Davidson:2000ne,  Rakshit:2004rj}. 
 %%%%%%%%%%%%%%%%%%%%%%%%%%%%%%%%%%%%%%%%%%%%%%%
\begin{figure}[H]
%\vspace{-0.45cm}
\centering
      \begin{subfigure}[t]{0.459\textwidth}
    \includegraphics[width=1\textwidth]{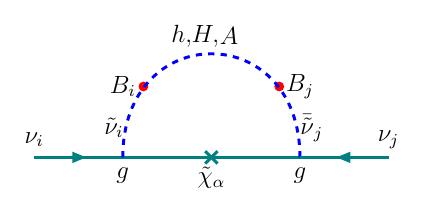}
    \caption{}
    \end{subfigure}
    \begin{subfigure}[t]{0.459\textwidth}
    \includegraphics[width=1\textwidth]{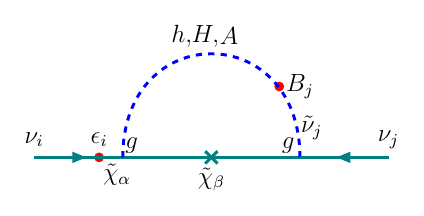}
    \caption{}
    \end{subfigure}
    \caption{Contribution to neutrino masses from one loop diagrams via - (a) $BB$ loop and (b)  $\epsilon B$ loop. For the $\epsilon B$ loop diagram there will be another diagram with $i \leftrightarrow j$. The cross represents mass insertion.}
    \label{fig:loop}
\end{figure}
%%%%%%%%%%%%%%%%%%%%%%%%%%%%%%%%%%%%%%%%%%%%%%%%%
Feynman diagrams of these two contributions are shown in Fig.~\ref{fig:loop}.  Fig.\ref{fig:loop}(a) corresponds to the $BB$ loop diagram where the blobs ($B_i$, $B_j$) represent the coupling between the sneutrinos and the neutral CP-even/odd Higgs bosons ($h,H,A$).  Fig~\ref{fig:loop}(b) corresponds to $\epsilon B$ loop where the mixing (generated by ${\epsilon_{i}}$)  between the neutrino and the neutralino appears in the external fermion line and the other blob ($B_j$) emerges on the internal scalar line. For the $\epsilon B$ loop diagram there will be another diagram with $i \leftrightarrow j$. In both the diagram Fig.~\ref{fig:loop}(a) \&  Fig.~\ref{fig:loop}(b), the cross on the neutralino line represents the Majorana mass insertion of neutralinos.
 %%%%%%%%%%%%%%%%

Combining all the contributions we can write neutrino mass matrix as \cite{Rakshit:2004rj,Barbier:2004ez}
\begin{equation} 
\begin{split}
\label{eq:total_mass}
[m_{\nu}]_{ij} & = [m_{\nu}]_{ij}^{\epsilon\epsilon} + [m_{\nu}]_{ij}^{BB} + [m_{\nu}]_{ij}^{\epsilon B} \\
& = X_T \epsilon_i \epsilon_j \sin^2\zeta + C_{ij} B_iB_j + (C_{ij}^\prime \epsilon_i B_j + i \leftrightarrow j) 
\end{split}
\end{equation}
Here $[m_{\nu}]_{ij}^{\epsilon\epsilon}$, $[m_{\nu}]_{ij}^{BB}$, $[m_{\nu}]_{ij}^{\epsilon B}$ correspond to the tree level contribution,  the $BB$ loop contribution and $\epsilon B$ loop contribution respectively.
Now if we look at the individual contribution, for the tree level, $\zeta$  represents the alignment between $\epsilon_i$ and $v_i$\cite{Chun:2002vp,Grossman:2000ex,Nardi:1996iy,Borzumati:1996hd}. For different basis choices, the alignment will be different\footnote{For more details on the basis choices corresponding to this alignment see \citep{Davidson:2000uc,Grossman:2003gq,Davidson:2000ne,Barbier:2004ez,Davidson:1996cc,Grossman:2000ex,Nardi:1996iy} }. In the tree level contribution,  $X_T$ is defined as \cite{Rakshit:2004rj,Grossman:2003gq}
\beq
\label{eq:XT}
X_T = {m_Z^2 m_{\tilde \gamma}\cos^2\beta \over 
\mu(m_Z^2 m_{\tilde \gamma}\sin 2\beta-M_1 M_2 \mu)}
\eeq 
where $m_{\tilde \gamma}\equiv \cos^2\theta_w M_1 + \sin^2\theta_w M_2$. In bRPV models, the neutral Higgses and the sneutrinos mix at the tree level via $B_i$ parameters. This leads to a finite mass splitting between the CP even and CP odd sneutrino mass eigenstates. This mass splitting of sneutrinos is responsible for the generation of Majorana neutrino mass at the one-loop level (see Fig.\ref{fig:loop}(a)). For a detailed discussion on the cancellation between different Higgs ($h,H,A$) mediated BB loop diagrams and the effect of sneutrino degeneracy from the $BB$ loop see Refs \cite{Grossman:2003gq,Rakshit:2004rj,Davidson:2000uc}. 
The $\epsilon B$ loop diagram involves both the bilinear  term $\epsilon_i$ and $B_j$. The combination of higgsino-neutrino mixing and Higgs-sneutrino mixing give rise to neutrino mass (see Fig.\ref{fig:loop}(b)). The $\epsilon B$ loop contribution is subleading to the $BB$ loop. Now if we consider that all the masses are at electroweak breaking scale ($\tilde{m}$), then the approximate contributions to the neutrino mass matrix (mentioned in Eq.\ref{eq:total_mass}) is given by \cite{Grossman:2003gq}: 
\begin{eqnarray}
\label{eq:tree}
[m_{\nu}]_{ij}^{\epsilon\epsilon}\sim{\cos^2\beta\over \tilde m} \epsilon_i \epsilon_j \sin^2\zeta \\
\label{eq:BB}
{[m_{\nu}]_{ij}^{BB} \sim \frac{g^2}{64\pi^2\cos^2\beta} \frac{B_iB_j}{\tilde{m}^3} ~\epsilon_H}\\ 
\label{eq:EB}
{ [m_{\nu}]_{ij}^{\epsilon B} \sim \frac{g^2}{64\pi^2\cos\beta} \frac{\epsilon_iB_j + \epsilon_jB_i}{\tilde{m}^2} ~\epsilon_H^\prime }
 \end{eqnarray}
where $\epsilon_H$ or $\epsilon_H^\prime$ arises due to the cancellation of the different Higgs ($h,H,A$) diagram in BB and $\epsilon$B loop respectively and depending upon the parameter space, they can suppress the mass contribution by several orders \cite{Rakshit:2004rj}. A recent article \cite{Dreiner:2022zsc} nicely summarises the different contributions to the light neutrino mass matrix for different model scenarios in the RPV context. The equations \ref{eq:tree} - \ref{eq:EB} are consistent with the ones quoted in the article with a small difference, that is the assumption of absence of sneutrino VEV ($v_i=0$) taken in \cite{Dreiner:2022zsc}.    

As already mentioned, only one of the three light neutrinos becomes massive at the tree level. If we consider the Normal Hierarchy (NH) scenario where the ordering of neutrino mass is $m_{\nu_3} > m_{\nu_2} > m_{\nu_1}$, then it is evident from  Eq.~\ref{eq:tree} that the heaviest neutrino mass is proportional to ${\cos^2\beta} (\epsilon_1^2 +  \epsilon_2^2 + \epsilon_3^2)$ i.e., tan$\beta$ acts as a suppression factor for $m_{\nu_3}$. The masses of the other two neutrinos, $m_{\nu_2}$ and  $m_{\nu_1}$, are generated at the one-loop level where the dominant contributions come from the $BB$ loop. In general, $B_i$ and $\epsilon_i$ are not related to each other and the leading contributions to $m_{\nu_2}$ comes from the $BB$ loop with an enhancement effect from the $ \frac{1}{\cos^2\beta}$  part as shown in Eq~\ref{eq:BB}. The lightest neutrino ($m_{\nu_1}$) in the NH scenario also can be massive from this same loop contribution. 
When the sneutrinos of different generations are non-degenerate, $m_{\nu_1}$ is proportional to the square of the sneutrino mass splitting between different generations. Hence to get the neutrino mass square differences and mixing angles in the existing ranges obtained from various neutrino oscillation experiments, we need tree level contribution as well as the loop contribution. This puts a restriction, among other parameters, on the choice of $\tan\beta$ which cannot be either very large or very small. It may be noted that for the Inverted Hierarchy (IH) scenario where $m_{\nu_2} > m_{\nu_1} > m_{\nu_3}$, the relations will change accordingly.  
%BB loops are also responsible for splitting of masses between the sneutrino CP-even and CP-odd states. The neutrino mass contributions at one loop level is proportional to this mass splitting \cite{Grossman:1998py, Davidson:2000uc, Davidson:2000ne, Grossman:1997is, Grossman:2003gq}. If the sneutrino mass splitting vanishes the neutrino mass arising from this kind of loop contribution also vanishes.

%\vspace{-0.99cm}

%%%%%%%%%%%%%%%%%%%%%%%%%%%%%%
\section{Computational Set-up and Numerical Constraints}
\label{sec:num}

In this section, we first summarize the numerical constraints that have been used in this study. The neutrino observables obtained from the latest global fit of different neutrino oscillation data, the most updated measurement of Higgs mass and its coupling strength in different decay modes along with low energy data from rare b-decays are considered in this analysis. Relevant limits derived from the LHC Run-I and Run-II data are also summarised in this section. Finally, we discuss the range of parameter space considered for scanning and summarize briefly the likelihood analysis implemented using the Markov Chain Monte Carlo (MCMC) algorithm.

%%%%%%%%%%%%%%%%%%%%%%%%%%%%%%
\subsection{Constraints from Neutrino observables}
\label{sec:obs_neut}
%%%%%%%%%%%%%%%%%%%%%%%%%%%%%%
Global analysis of neutrino oscillation data provides us with two mass-square differences and three mixing angles. The mass-squared differences are defined as 

\begin{equation}
\begin{split}
\Delta m^2_{21} &= m^2_2 - m^2_1 \nonumber \\
|\Delta m^2_{31}| &= |m^2_3 - m^2_1| \nonumber
\end{split}
\end{equation}

where, the $m_i~(i=1,~2,~3)$ represent the physical masses of the three light neutrinos. The sign of $\Delta m^2_{31}$ (or $\Delta m^2_{32}$) remains unknown to date, which gives rise to the two hierarchial (NH and IH) scenarios. The three relevant mixing angles between different generations are represented by $\theta_{12}$, $\theta_{13}$ and $\theta_{23}$.  
The mixing angles can be calculated from the $3\times 3$ light neutrino mass matrix, also known as PMNS matrix \cite{Donini:1999jc,Akhmedov:1999uz,Giganti:2017fhf} which looks as shown in equation \ref{eq:pmns_matrix}.  
\begin{equation}
\label{eq:pmns_matrix}
\left( \begin{matrix}
  c_{12}c_{13} & s_{12}c_{13} & s_{13} \\
  -s_{12}c_{23} - c_{12}s_{13}s_{23} & c_{12}c_{23} - s_{12}s_{13}s_{23} & c_{13}s_{23} \\ 
  s_{12}s_{23} - c_{12}s_{13}c_{23} & -c_{12}s_{23} - s_{12}s_{13}c_{23} & c_{13}c_{23} 
\end{matrix} \right )
\end{equation} 
Here, $c_{ij}$ and $s_{ij}$ represent $\cos\theta_{ij}$ and $\sin\theta_{ij}$ ($i$, $j$ are generation indices) respectively. 
For simplicity, we have kept the CP-violating phase as zero and work with the five aforementioned observables, namely, $\Delta m_{21}^2$, $|\Delta m_{31}^2|$, $\theta_{12}$, $\theta_{13}$ and $\theta_{23}$. Several groups have done global fits with neutrino oscillation data obtained from different experiments \cite{deSalas:2020pgw, Esteban:2020cvm}. For this analysis, we have used the best-fit values and 
%% have used or we use !! keep the same format
 1$\sigma$ ranges of neutrino oscillation parameters obtained by the updated global fit 
 in Ref. \cite{deSalas:2020pgw}. The best-fit points along with 1$\sigma$ range  
of these parameters in NH and IH scenarios are summarized in 
Table~\ref{tab:neutrino_obs}. 
%%%%%%%%%%%%%%%%%%%%%%%%%%%%%%%%%%%%%%%%%
\begin{table}[h!]
\centering
\begin{tabular}{||l|l||} 
    \hline\hline
 {\bf Observable} & {\bf Best-fit value $\pm$1$\sigma$} \\
	\hline\hline
	$\Delta m^2_{21}$[$10^{-5}$eV$^2$] & \hspace{7mm}7.50$^{+0.22}_{-0.20}$ \\
	\hline
	$|\Delta m^2_{31}|$[$10^{-3}$eV$^2$][NH] & \hspace{7mm}2.55$^{+0.02}_{-0.03}$ \\
	$|\Delta m^2_{31}|$[$10^{-3}$eV$^2$][IH] & \hspace{7mm}2.45$^{+0.02}_{-0.03}$ \\
	\hline
	\hspace{10mm}$\theta_{12}/^{\circ}$ & \hspace{7mm}34.3 $\pm$ 1.0 \\
	\hline
	\hspace{5mm}$\theta_{13}/^{\circ}$[NH]& \hspace{7mm}8.53$^{+0.13}_{-0.12}$  \\
	\hspace{5mm}$\theta_{13}/^{\circ}$[IH]& \hspace{7mm}8.58$^{+0.12}_{-0.14}$  \\
	\hline
	\hspace{5mm}$\theta_{23}/^{\circ}$[NH] & \hspace{7mm}49.26 $\pm$ 0.79 \\
	\hspace{5mm}$\theta_{23}/^{\circ}$[IH] & \hspace{7mm}49.46$^{+0.60}_{-0.97}$ \\
    \hline\hline
\end{tabular}
\caption{
%List of neutrino sector observables provided by the updated global fit analysis \cite{deSalas:2020pgw} of neutrino oscillation data obtained from different neutrino experiments. Here NH refers to Normal Hierarchy scenario and IH refers to Inverted Hierarchy scenario. Here mixing angles are shown in degrees.
List of neutrino sector observables provided by the updated global fit analysis \cite{deSalas:2020pgw} of neutrino oscillation data obtained from different neutrino experiments. Here NH and IH refers to Normal Hierarchy and Inverted Hierarchy scenario.}
\label{tab:neutrino_obs}
\end{table} 
%%%%%%%%%%%%%%%%%%%%%%%%%%%%%%%%%%%%%%%%
We have considered both normal and inverted hierarchy scenarios separately in our analysis for comparative study. The choice of hierarchy is expected to be reflected in the resulting parameter space. Note that, the neutrino oscillation experiments also measure the CP-violating phase ($\delta_{\rm CP}$). The updated global analysis quotes the best-fit values (in degrees) as $194^{+24}_{-22}$ for NH and $284^{+26}_{-28}$ for IH scenario respectively \cite{deSalas:2020pgw}. The $\delta_{\rm CP}$ corresponding to NH scenario is quite consistent with $\pi$ well within $1\sigma$, which is consistent with zero CP-violation in the neutrino sector. The existing uncertainties on this parameter are also much bigger compared to the mixing angles and mass-squared differences. Adding the $\delta_{\rm CP}$ parameter does not restrict the parameter space any further and hence we choose to keep $\delta_{\rm CP}=0$ throughout our analysis.   
%%%%%%%%%%%%%%%%%%%%%%%%%%%%%%%%%%%%%%%%%%%%%%%%%%%%%%%

\subsection{Constraints from Collider experiments}
\label{sec:constraint}
%%%%%%%%%%%%%%%%%%%%%%%%%%%%%%%%%%%%%%%%%%%%%%%%%%%%%%%
While fitting the neutrino physics observables, one also needs to ensure that the fitted model particle spectra obey other experimental constraints. The bRPV term gives rise to mixing among the sneutrino and Higgs states of the MSSM whereas the charged sleptons now mix with the charged Higgs sectors. 
In addition to that, to fit the neutrino data we are also varying 
$\mu$ and $\tan\beta$, which impacts the mass of the lightest CP even 
Higgs boson\footnote{It is consistent with the SM like 125 GeV Higgs boson observed by LHC collaborations\cite{ATLAS:2012yve,CMS:2012qbp}.}and its coupling strengths with SM particles. These are quite precisely measured and as a result, restrict the choices of the parameters affecting them. Apart from the neutrino and Higgs sector, we also consider constraints arising from the branching of rare B-hadron decays such as $BR(B \rightarrow X_s \gamma)$ and $BR(B_s \rightarrow \mu^+  \mu^-)$. Last but not least, the existing constraints on SUSY particles from various direct searches are also duly taken into account.

%%%%%%%%%%%%%%%%%%%%%%%%%%%%%%
\subsubsection{Constraints from Higgs Sector}
\label{subsubsec:higgs_mass}
%%%%%%%%%%%%%%%%%%%%%%%%%%%%%%
The measured mass of SM-like Higgs boson obtained from the combined data of the ATLAS and CMS experiments is 125.09 $\pm$ 0.21(stat.) $\pm$ 0.11(syst.) GeV\cite{ATLAS:2015yey}. Taking into account the theoretical uncertainty of the Higgs mass calculation within the SUSY framework, we consider $\pm$3 GeV window for Higgs mass around the best-fit value \cite{Allanach:2004rh}. Apart from the mass, the signal strengths of 125 GeV Higgs are also precisely measured by both the CMS and ATLAS collaborations \cite{cms_web1, ATLAS:2021vrm}. The updated results of the coupling strength modifiers  ($\kappa_i$), i.e., BSM over SM ratios of the coupling strengths for a particular decay mode $i$,  along with their 1$\sigma$ uncertainties obtained by CMS collaboration from LHC Run-II data with luminosity $\mathcal{L}$ = 137 $fb^{-1}$ are summarized in Table \ref{tab:higgs_coupling}. 
\begin{table}[h!]
	\centering
	\begin{tabular}{||c|c||}
	\hline\hline
	\multicolumn{1}{||c|}{\bf{Coupling Strength}} &
	\multicolumn{1}{c||}{\bf{Best-fit$\pm$ 1$\sigma$}} \\
	\hline\hline
	\hspace{5mm} $\kappa_z$ & 0.96 $\pm$ 0.07\\
	\hline
	\hspace{5mm} $\kappa_w$ & 1.11$^{+0.14}_{-0.09}$ \\
	\hline
	\hspace{5mm} $\kappa_b$ & 1.18$^{+0.19}_{-0.27}$ \\
	\hline
	\hspace{5mm} $\kappa_t$ & 1.01 $\pm$ 0.11 \\
	\hline
	\hspace{5mm} $\kappa_{\mu}$ & 0.92$^{ +0.55}_{ -0.87}$ \\
	\hline
	\hspace{5mm} $\kappa_{\tau}$ & 0.94 $\pm$ 0.12 \\
	\hline
	\hspace{5mm} $\kappa_{\gamma}$ & 1.01$^{+0.09}_{-0.14}$ \\
%	\hline
%	\hspace{5mm} \tcm{$\kappa_g$}!!! delete/move to text & 1.16$^{+0.12}_{-0.11}$ \\
	\hline\hline
	\end{tabular}
	\caption{Higgs boson coupling strength modifiers obtained by the CMS collaboration using LHC Run-II 137 $fb^{-1}$ data \cite{cms_web1}. }
	\label{tab:higgs_coupling}
\end{table}

Note that, although some of the data points have very similar best-fit and uncertainty ranges (e.g., $\kappa_z$, $\kappa_{\tau}$), there are considerable differences in some other measurements (e.g., $\kappa_b$, $\kappa_{\mu}$) obtained from the ATLAS collaboration~\cite{ATLAS:2021vrm}. These differences especially in the best-fit points with similar uncertainty can slightly change the favored parameter space. Hence, we cross-check our results using ATLAS data \cite{ATLAS:2021vrm} as well and comment on how much change is expected. The coupling strengths in the present model framework have been computed using SPheno which does a full two-loop calculation for the Higgs sector. Note that the CMS collaboration also quotes their measurement of the Higgs-gluon-gluon effective coupling strength as $\kappa_g$ = 1.16$^{+0.12}_{-0.11}$, which we have not included in our analysis. This is because the effective Higgs-gluon-gluon coupling strength is quite sensitive to the choice of the SUSY breaking scale. Hence one would in principle have to vary the scale as well as a parameter, which affects among other things, the 125 GeV Higgs mass itself. That in turn prevents us from fixing some Higgs sector parameters, such as $A_t$. To reduce the number of input parameters and thereby computation time and since our main focus remains on the neutrino sector, we avoid this scenario.
% However, \tcb{we do comment on how our result is impacted if we add $\kappa_g$ as an observable keeping number of input parameters same. .. yet not added !! in appendix ? \textbf{DELETE}}

%\tcc{ For numerical calculations, we have used SPheno \cite{Porod:2003um,Porod:2011nf} which calculates the Higgs masses considering up to two-loop correction \cite{Goodsell:2014bna} and all the other particle masses at one-loop level. The model was implemented in SPheno using SARAH \cite{Staub:2008uz,Staub:2010jh,Staub:2015kfa}.}

%%%%%%%%%%%%%%%%%%%%%%%%%%%%%%%%%%%%%%%%%%%%
\subsubsection{Constraints from Flavor Physics}
\label{subsubsec:flavor_physics}
%%%%%%%%%%%%%%%%%%%%%%%%%%%%%%%%%%%%%%%%%%%%
Low energy flavor observables play an important role in constraining the SUSY parameter space. Branching ratios (BR) of flavor changing neutral current (FCNC) \cite{Archilli:2017xmu} decays like $B \rightarrow X_s  \gamma$ and $B_s \rightarrow \mu^+ \mu^-$, can put some non-trivial bounds on the MSSM parameter space. The world average of BR($B \rightarrow X_s  \gamma$) at present is (3.32 $\pm$ 0.15)$\times 10^4$ \cite{HFLAV:2019otj}. For BR($B_s \rightarrow \mu^+  \mu^-$), we have considered the range (3.09$^{+0.46 + 0.15}_{-0.43 - 0.11}$)$ \times 10^{-9}$ provided by the LHCb collaboration \cite{LHCb:2021vsc} after combined analysis of data collected with center of mass energies $\sqrt{s}$ = 7, 8, and 13 TeV. Adding the errors in quadrature, we have used BR$(B_s \rightarrow \mu^+  \mu^-)$ = (3.09$^{+0.48}_{-0.44})\times 10^{-9}$ in our analysis. The calculations for these branching ratios are performed using FlavorKit \cite{Porod:2014xia} which is integreated within SPheno \cite{Porod:2003um,Porod:2011nf} through SARAH \cite{Staub:2008uz,Staub:2010jh,Staub:2015kfa}. 
%%%%%%%%%%%%%%%%%%%%%%%%%%%%%%%%%%%%%%%%%%%%

\subsubsection{Constraints from direct searches of sparticles from the LHC:}
\label{subsec:susy_limit}
%%%%%%%%%%%%%%%%%%%%%%%%%%%%%%%%%%%%%%%%%%%%

The LHC collaboration has extensively searched for the supersymmetric partners of the SM particles (sparticles) from Run-I and Run-II data in various final states and without any statistically significant deviation of data over the SM prediction, the LHC has 
imposed stringent lower limits on the sparticle masses. For a summary of the ATLAS and CMS SUSY searches see Ref.\cite{atlas_web,cms_web}. Here we briefly mention the most stringent bounds which are valid for simplified scenarios with specific assumptions on 
branching ratios and mostly for massless or relatively light neutralino. While doing so, we summarise the limits corresponding to both R-parity conserving and violating scenarios. Unless the R-parity violating couplings are large enough, in some cases the bounds corresponding to the R-parity conserving scenario may also be applicable. Eventually, the applicable limits depend on the mass and decay branching ratios of the relevant particle. We have applied the limits accordingly. 
\begin{itemize}
\item In the simplified RPC-SUSY framework with different choices of decay mode and branching, the ATLAS and CMS collaborations have now pushed the lower limit of gluino mass to $\sim$ 2.0 - 2.3 TeV for $\mlspone$ upto 600  GeV~\cite{CMS:2019zmd,ATLAS:2021twp,CMS:2021beq,ATLAS:2020syg}. The LHC  has also pushed the light squarks mass to $\sim$ 1.85 TeV \cite{CMS:2019zmd,ATLAS:2020syg,ATLAS:2022zwa}, the lightest stop to $\sim$ 1.0 - 1.3 TeV \cite{ATLAS:2020dsf,ATLAS:2020xzu,ATLAS:2021hza,CMS:2021beq,CMS:2019ybf} for $\mlspone$ upto 300  GeV. 
In RPV SUSY scenarios if LSP decays to charged leptons via LLE couplings, then the limits obtained by the LHC are relatively stronger. For example, the ATLAS collaboration has excluded gluino mass up to 2.5 TeV \footnote{For UDD scenarios with gluino cascade decay as $\widetilde{g} \to (q \bar{q}) \lspone \to (q \bar{q}) q q q$,  $m_{\widetilde{g}}$ between 1 - 1.85 TeV are excluded at 95\% CL depending on  $\mlspone$ \cite{ATLAS:2018umm}.} in such scenarios \cite {ATLAS:2021yyr}.  We have kept the squarks and gluino masses at 3 TeV to evade the current  LHC constraints. 
\item The limits on electroweak (EW) sparticles i.e., sleptons and electroweakinos are relatively weaker compared to strong sparticles. 
For example, in RPC scenarios, the LHC collaboration has searched for electroweakinos for different decay modes like slepton mediated, WZ and Wh mediated final states and has excluded wino like ${\tilde{\chi}_1^{\pm}}$ upto $\sim$ (1.0 - 1.4) TeV \cite{ATLAS:2019lff,ATLAS:2021yqv,ATLAS:2022zwa,CMS:2021cox}. 
For slepton pair production, slepton mass upto $\sim$ 700 GeV is excluded for massless neutralino \cite{ATLAS:2019lff} for the universal slepton mass scenario. However, it should be noted that these strong limits are not 
always applicable to the overall parameter space of the realistic SUSY scenarios, e.g.,  compressed SUSY scenarios \citep{ATLAS:2019lng}.  
%For pure Higgsino pair production, the light electroweak sparticles around 250 GeV are still allowed.}
The ATLAS and CMS collaborations have also interpreted the limits in RPV SUSY scenarios with $LLE$, $UDD$ and bRPV couplings. For $LLE$ type coupling, slepton and chargino (wino type) masses are excluded upto 1.2 and 1.6 TeV respectively \cite{ATLAS:2021yyr}. Limits in models with  $UDD$ coupling get drastically reduced as compared to RPC scenarios \cite{Barman:2020azo}. Again for pure higgsino type $\tilde{\chi}_1^{\pm}\tilde{\chi}_2^0$ pair production in RPC scenarios, the lower bound on $m_{\tilde{\chi}_1^{\pm}=\tilde{\chi}_2^0}$ is reduced to $\sim$ 210 GeV \cite{ATLAS:2021moa}. On the other hand, higgsinos in bRPV scenarios are excluded upto 440 GeV  \cite{ATLAS:2023lfr}. 
It is worth mentioning that a recent article \cite{Dreiner:2023bvs} has provided an updated detailed summary of the possible gaps in RPV-MSSM searches at the LHC. In the Refs.~\cite{Dercks:2017lfq, Dreiner:2023bvs}, the authors have meticulously classified the various possible trilinear RPV-MSSM signatures at the LHC. In their analysis, they have studied both direct and indirect production of various LSPs and derive limits on SUSY masses, which are comparable or an improvement on those obtained in the R-parity conserving scenarios. However, these limits are not directly applicable in our bilinear RPV scenario. For more details, refer to \cite{Dreiner:2023bvs}.

\item The most stringent limit on $M_A$ comes from the heavy Higgs searches in the  $H/A \rightarrow \tau^+ \tau^-$ decay channel and  
typically $M_A <$ 1.5 (1.0) TeV is excluded for $\tan\beta <$ 21 (8).
\cite {CMS:2018rmh,ATLAS:2020zms}. 

\end{itemize}

\subsection{Survey of parameter space }
\label{sec:parameter}
%%%%%%%%%%%%%%%%%%%%%%%%%%%%%%
In the bRPV model, we have nine RPV parameters - three bRPV couplings ($\epsilon_1$, $\epsilon_2$, $\epsilon_3$), three corresponding soft coupling parameters ($B_1$, $B_2$, $B_3$), and three sneutrino VEV parameters ($v_1$, $v_2$, $v_3$). 
The resulting light neutrino masses are quite sensitive to the choices of all these nine parameters. Apart from that, we also have different MSSM parameters which are essential to achieve our objective. Among them, $\mu$ and $\tan\beta$ are the most relevant ones. 
Given the large number of independent parameters in low scale MSSM, we fix some parameters which are not directly affecting the neutrino sector. For example, trilinear coupling for the third generation squark ($A_t$) is a very important parameter to achieve a 125 GeV Higgs in the model but it does not have a direct impact on the light neutrino masses and mixing angles. 
As large values of $A_t$ is required to obtain $m_h \sim$ 125 GeV, we have chosen  
$A_t$ = -3.5 TeV. We have also fixed all the three generation squarks soft masses ($m_{\tilde{q}}$) and slepton soft masses ($m_{\tilde{l}}$)  at 3 and 2 TeV respectively to ensure that none of the current exclusion bounds on sparticles masses affect our parameter space. Similarly, we have fixed the bino ($M_1$), wino ($M_2$), gluino ($M_3$) soft masses and $M_A$ at 0.3, 1.2, 3.0  and 3.0 TeV respectively.

%Hence we keep it fixed at a reasonable numerical value (-3.5 TeV)  so that we get the mass for lighter CP even neutral Higgs ($m_h$)  around 125 GeV. 
%Similarly, we keep  $M_A$ at 3 TeV **(Justification \tcb{mention about the limits in MA-TB plane}), bino, wino and gluino soft masses at 300 GeV, 1.2 TeV, and 3 TeV respectively. 

Following a literature survey \citep{Barbier:2004ez, Diaz:2014jta} and some preliminary computation, we decided on exhaustive ranges for the bRPV model parameters $\epsilon_i$, $v_i$ and $B_i$ to ensure that the light neutrino mass square differences and mixing angles are generated in the correct order. Since our objective is to probe both the normal and inverted hierarchy scenarios, we do not presume any hierarchy in the choices of these parameters generation-wise and keep the ranges uniform over all three generations. We keep the choices conservative for the other two input parameters $\mu$ and $\tan\beta$. The ranges of these input parameters are enlisted in Table \ref{tab:parameter_range}. 
\begin{table}[h!]
	\centering
	\begin{tabular}{||l|l|l||}
    \hline\hline
	{\bf Input Parameters} & {\bf Lower Range} & {\bf Upper Range} \\
	\hline\hline
	$\mu$ & \hspace{2mm} 1 TeV & \hspace{2mm} 3 TeV\\
	\hline
	$\tan\beta$ & \hspace{5mm} 1 & \hspace{4mm} 60\\
	\hline
	$\epsilon_i (i = 1,2,3)$ & \hspace{1mm} -1.0 GeV & \hspace{2mm} 1.0 GeV\\
	\hline
	$v_i (i = 1,2,3)$ & \hspace{1mm} $10^{-8}$ GeV & \hspace{2mm} 0.1 GeV\\
	\hline
	$B_i (i = 1,2,3)$ & \hspace{1mm} $10^{-3}$ GeV & \hspace{2mm} 10 TeV \\
	\hline\hline
	\end{tabular}
	\caption{Ranges of eleven input parameters considered in our analysis.}
	\label{tab:parameter_range}
\end{table}
%%%%%%%%%%%%%%%%%%%%%%%%%%%%%%%%%%%%%%
%%%%%%%%%%%%%%%%%%%%%%%%%%%%%%%%%%
\subsection{Analysis set-up}
\label{sec:analysis}
%%%%%%%%%%%%%%%%%%%%%%%%%%%%%%%%%%

Given the data and set of model parameters, as described earlier, we now proceed to calculate the posterior probability distribution in order to locate the favored parameter space. This is obtained using the MCMC technique which maximizes the likelihood function (or minimizes $\chi^2$) defined as
\begin{equation}
    L \propto \exp(-\mathcal{L})
\end{equation}
where $\mathcal{L}$ is the negative of the log-likelihood and calculated using
\begin{equation}
    \mathcal{L} = \frac{\chi^2}{2} = \frac{1}{2}\mathlarger{\mathlarger{\sum}}_{i=1}^{n_{\rm obs}}\left[ \frac{\Gamma_i^{\rm obs}-\Gamma_i^{\rm th}}{\sigma_i} \right]^2
\end{equation}
where $\Gamma_i^{\rm obs}$ represents the set of $n_{\rm obs}$ observed data points with corresponding errors $\sigma_i$ on them and $\Gamma_i^{\rm th}$ is the calculated value of each observable using our theoretical model. Altogether, we have total 15 independent observables (two neutrino mass-squared differences and three mixing angles, SM-like Higgs mass and seven coupling modifiers, two flavor constraints from rare b-decay) and 11 free parameters ($\epsilon_i$, $B_i$, $v_i$, $\tan\beta$ and $\mu$), so that the degrees of freedom (DoF) for the $\chi^2$ distribution is 4. 

%\tcc{ For numerical calculations, we have used SPheno \cite{Porod:2003um,Porod:2011nf} which calculates the Higgs masses considering up to two-loop correction \cite{Goodsell:2014bna} and all the other particle masses at one-loop level. The model was implemented in SPheno using SARAH \cite{Staub:2008uz,Staub:2010jh,Staub:2015kfa}.}
The bRPV SUSY spectrum is generated by SPheno \cite{Porod:2003um,Porod:2011nf} which calculates the Higgs masses considering up to two-loop correction \cite{Goodsell:2014bna} and all the other particle masses at one-loop level. The bRPV model was implemented in SPheno using SARAH \cite{Staub:2008uz,Staub:2010jh,Staub:2015kfa}.
For the MCMC-based likelihood analysis, we use publicly available code
\textit{emcee} \cite{Foreman-Mackey_2013} which is a Python implementation of the affine-invariant ensemble sampler. The code itself ensures an efficient exploration of parameter space even if there are strong degeneracies among those. We use a flat prior on all the parameters as mentioned in Table~\ref{tab:parameter_range}. To get the desirable acceptance fraction for the proposed MCMC steps, we use a relatively higher number of random walkers (500) each with a sufficient number of steps (chain length). An auto-correlation analysis has also been carried out which ensures that the convergence criterion for each chain is well-satisfied. 

%We identify 1$\sigma$ and 2$\sigma$ regions of different correlated parameters. \tcb{For any two dimensional parameter space under consideration, the ellipse contour will be centered \tcm{!!!!! FOR IH it's not} at the best-fit point of that parameter space where log-likelihood is maximum and $\chi^2$ is minimum ($\chi^2_m$). Then we find out the points with $\chi^2$ value as $\chi^2$ = $\chi^2_{m} +2.30$ for 1$\sigma$ contour. Similarly for 2$\sigma$ region it will be $\chi^2$ = $\chi^2_{m} + 6.18$.} \tcm{taken from where ??} Lastly we discuss the results we got from our analysis. 

%%%%%%%%%%%%%%%%%%%%%%%%%%%%%%%%%%
\section{Results and discussion}
\label{sec:result}
%%%%%%%%%%%%%%%%%%%%%%%%%%%%%%%%%%
In this section, we present our results for both the hierarchy scenarios - NH and IH. For each scenario, we present the marginalized distributions for different input 
parameters. We also mention the best-fit and mean values of the parameters along with the 
$\chi^2_{min}$. Finally, we compare the allowed parameter space in NH and IH scenarios.  
%%%%%%%%%%%%%%%%%%%%%%%%%%%%%%%%

\subsection{Normal Hierarchy scenario}
\label{subsec:nh}  
%%%%%%%%%%%%%%%%%%%%%%%%%%%%%%%%

In the NH scenario, the heaviest of the three light neutrinos is dominantly $\tau$ flavored. 
The second heaviest state is a nearly equal admixture of all three flavors 
($e, \mu, \tau$) whereas the lightest neutrino mass eigenstate is dominantly $e$ flavored. 
 This hierarchy is expected to be highlighted by the choices of the neutrino sector parameters in the most probable region. 
 %% \sout{We present our results in the form of 1$\sigma$ and 2$\sigma$ allowed regions obtained from the marginalized 2-D distributions on various input parameter planes... shifted to next para.} %\textcolor{red}{Write a couple of lines how the 1$\sigma$ and 2$\sigma$ regions are obtained.} 

%%%%%%%%%%%%%%%%%%%%%%%%%%%%%%%%%%%%%%%%%%%%%%%%%%%%%%%%
\begin{table}[!htb]
    \centering
	\begin{tabular}{||l|l|l||l|l||}
	\hline\hline
	 \multicolumn{3}{||c||}{Input parameters} & \multicolumn{2}{c||}{Output observables}	\\
	\hline
    Para- & Best-fit & \hspace{2mm} Mean value [95$\%$C.L.]  & Observable & Best-fit \\ 
    meter & value &  & & value \\ 
	\hline\hline
	$\epsilon_1$ & -0.0072 & -0.0045 [-0.0183, 0.0054] & $\Delta m_{21}^2$[eV$^2$] & 7.51$\times10^{-5}$  \\
	\hline
	$\epsilon_2$ & -0.0160 & -0.0089 [-0.0218, 0.0034] & $\Delta m_{31}^2$[eV$^2$] & 2.55$\times10^{-3}$ \\
	\hline
	$\epsilon_3$ & -0.0279 & -0.0311 [-0.0487, -0.0068] & $\theta_{13}$ & 8.51$^{\circ}$  \\
	\hline
	$v_1$  & 0.00038 & 0.00034 [0.00019, 0.00055] & $\theta_{12}$ & 34.04$^{\circ}$ \\
	\hline
	$v_2$ & 0.00052 & 0.00040 [0.00022, 0.00060] & $\theta_{23}$ & 49.29$^{\circ}$ \\
	\hline
	$v_3$ & 0.00091 & 0.00097 [0.00061, 0.00122] & $m_h$ [GeV] & 124.61\\
	\hline
	$B_1$ & 461.78 & 548.38~ [264.50, 791.66] & $BR(B \rightarrow X_s \gamma)$ & 3.14$\times10^{-4}$  \\
	\hline 
	$B_2$ & 198.62 & 276.03 ~[5.01, 515.16] & $BR(B_s \rightarrow \mu^+ \mu^-)$ & 3.21$\times10^{-9}$  \\
	\hline
	$B_3$ & 1760.66 & 1917.76 [1359.46, 2355.98] & $k_z$ & 1.0  \\
	\hline
	$\mu$ & 1293.80 & 1249.33 [1028.70, 1449.71] & $k_w$ & 1.0  \\
	\hline
	$\tan\beta$ & 12.11 & 12.81 ~~~[8.76, 15.67] & $k_b$ & 1.001854 \\
	\hline
	&  &  & $k_{\tau}$ & 1.001854 \\
	\hline
	 &  &  &  $k_{\mu}$ & 1.001854 \\
	\hline
	 &  &  &  $k_t$ & 0.9999874  \\
	\hline
	 &  &  & $k_{\gamma}$ & 1.075042 \\
	\hline\hline
	\multicolumn{5}{|c|}{ $\chi^2_{min} = $ 3.46~~~ ~~~~~ ${\chi^2_{min}}/${DoF} = 0.865} \\
%	\hline\hline
%	\multicolumn{5}{|c|}{$\frac{\chi^2_{min}}{DoF}$ = $\frac{3.46}{4}$ = 0.865} \\
	\hline\hline
		\end{tabular} 
	\caption{Best-fit, mean values along with the 95$\% C.L.$ of all the free parameters and best-fit values for the observables in NH scenario are shown here. The last row represents $\chi^2_{min}$ and ${\chi^2_{min}}/${DoF}.}
	\label{tab:nh_input_output}
\end{table}

The best-fit, mean values along with 95\% $C.L.$ of input parameters are listed in the 
Table \ref{tab:nh_input_output}. The observable values for the best-fit point are also mentioned in Table \ref{tab:nh_input_output}. 
The best-fit point for NH corresponds to $\chi^2_{min}$ = 3.46,  
degrees of freedom (DoF) = 4 and ${\chi^2_{min}}/${DoF}  = 0.865\footnote{ We have also checked that one gets similar parameter space with the ATLAS data~\cite{ATLAS:2021vrm}. } 
%Considering the ATLAS Higgs coupling data (Table 8 of Ref.\citep{ATLAS:2021vrm}), our analysis gives $\chi^2_{min}$ = 6.4,    and ${\chi^2_{min}}/${DoF}  = 1.6. }
%It may be noted that the contribution to $\chi^2$ is large from Higgs couplings as compared to the CMS data.}} 

%For NH scenario using both the ATLAS and CMS results (total 14 data points for coupling strength),  our setup gives $\chi^2_{min} \sim$  8.9,  ${\chi^2_{min}}/{DoF} $ = 8.9/11 = 0.81.

The contribution to the $\chi^2_{min}$ from the neutrino observables for the best-fit point is~$\sim$ 0.097. It is evident that the model can fit the neutrino oscillation data quite nicely. The neutrino masses at the above-mentioned best-fit point are: $m_{\nu_1} = 3.58\times10^{-6}$ eV, $m_{\nu_2} = 8.67\times10^{-3}$ eV and $m_{\nu_3} = 5.05\times10^{-2}$ eV and their sum is %$\sum_{i=1}^{3} m_{\nu_i} 
$\sum m_{\nu_i} 
= 0.059$ eV. It may be noted that the 2$\sigma$ upper limit on the sum of neutrino masses, in NH scenario, coming from the cosmological data \cite{deSalas:2020pgw} is $\sum m_{\nu_i} < 0.12$ eV.

%%%%%%%%%%%%%%%%%%%%%%%%%%%%%%%%%%%%%%%%%%%%%%%%%%%%%%%%
\begin{figure}[!htb]
\begin{center}
\includegraphics[angle =360, width=0.8\textwidth]{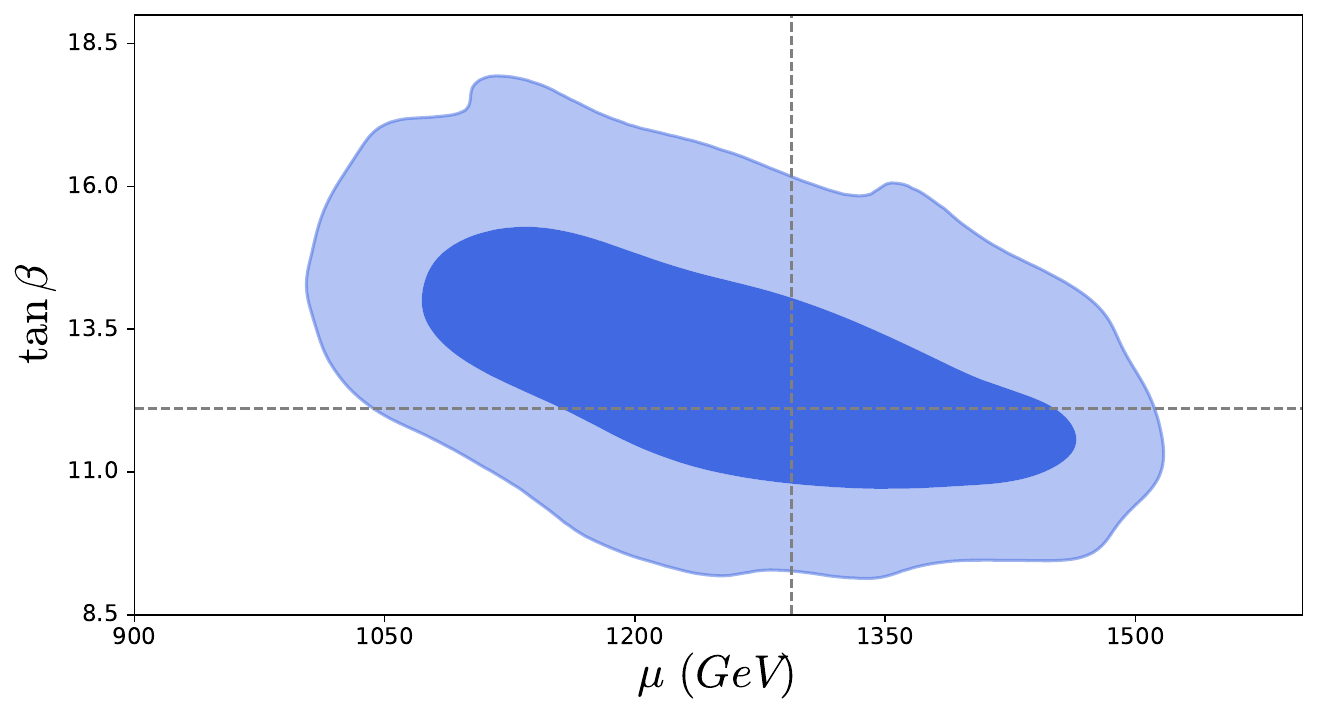}
\caption{Marginalized posterior distribution with 68\% (dark blue) and 95\% (light blue) $C.L.$ contours in the $\mu -\tan\beta$ plane for NH scenario. The dashed grey lines indicate the best-fit values.}
    \label{fig:nh_mu_tanbeta}
\end{center}
\end{figure}
%%%%%%%%%%%%%%%%%%%%%%%%%%%%%%%%%%%%%%%%%%%%%%%%%%%%%%%%

%%%%%%%%%%%%%%%%%%%%%%%%%%%%%%%%%%%%%%%%%%%%%%%%%%%%%%%%
\begin{figure}[!htpb]
\begin{center}
\includegraphics[angle =360, width=0.75\textwidth]{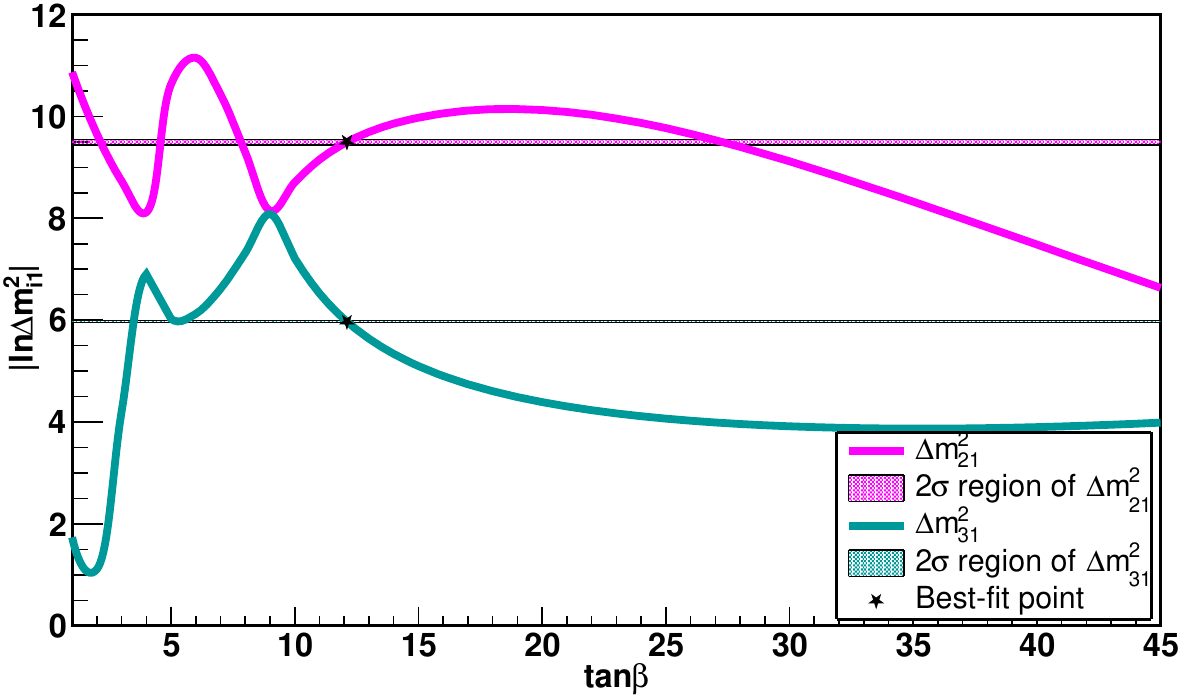}
\caption{The mass square differences ($\Delta m^2_{21}$ and 
$\Delta m^2_{31}$) vs $\tan\beta$ plot where all other parameters are kept fixed at the best-fit point. The line with magenta (cyan) color corresponds to $\Delta m_{21}^2$ ($\Delta m_{31}^2$) and the horizontal magenta (cyan) shaded region shows the corresponding  2$\sigma$ allowed region obtained from global analysis of neutrino oscillation data~\cite{deSalas:2020pgw}. The mass-squared differences are taken in $eV^2$ units.}
    \label{fig:mass_split}
\end{center}
\end{figure}
%%%%%%%%%%%%%%%%%%%%%%%%%%%%%%%%%%%%%%%%%%%%%%%%%%%%%%%%

\begin{figure}[!htpb]
    \begin{subfigure}[t]{0.49\textwidth}
    \includegraphics[width=1\textwidth]{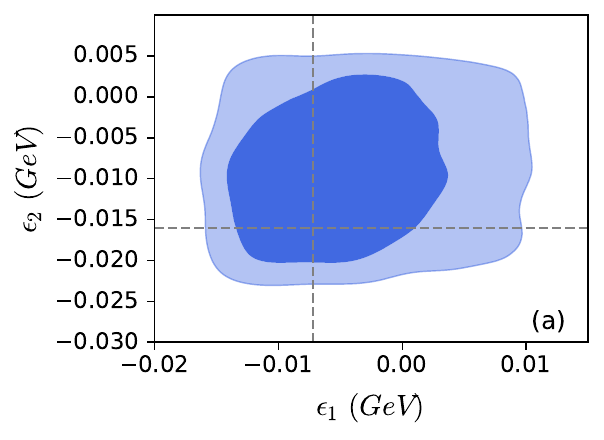}
   % \caption{}
    \label{fig:nh_eps12}
    \end{subfigure}
    \begin{subfigure}[t]{0.49\textwidth}
    \includegraphics[width=1\textwidth]{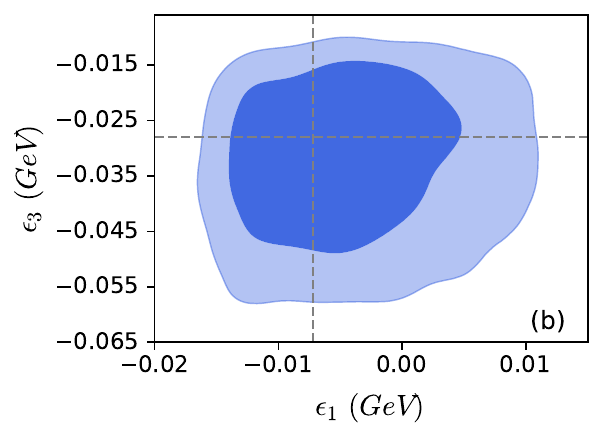}
    \end{subfigure}
    \\[\smallskipamount]
    \begin{subfigure}[t]{0.49\textwidth}
    \includegraphics[width=1\textwidth]{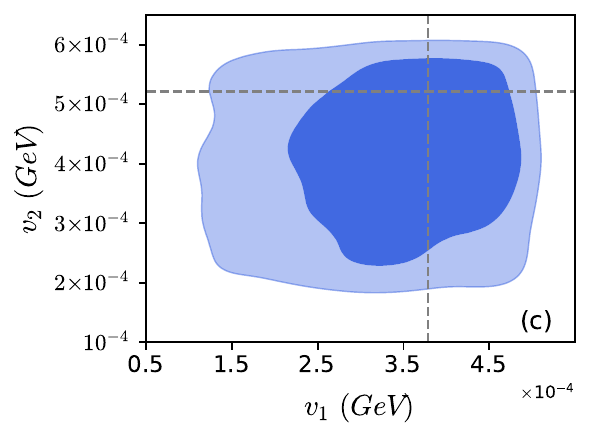}
    \end{subfigure}
    \hfill
    \begin{subfigure}[t]{0.49\textwidth}
    \includegraphics[width=1\textwidth]{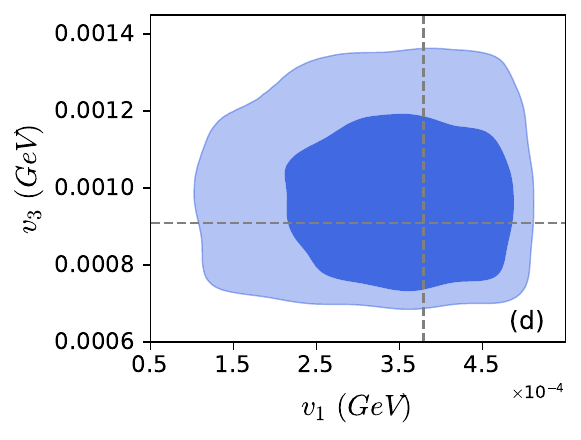}
    \end{subfigure}
    \\[\smallskipamount]
    \begin{subfigure}[t]{0.49\textwidth}
    \includegraphics[width=1\textwidth]{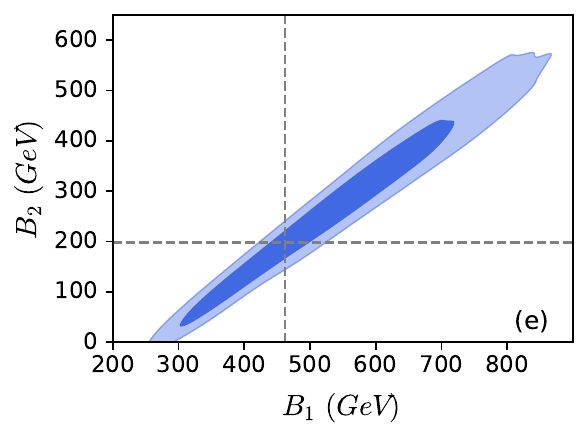}
    %\caption{}
    \label{fig:nh_B12}
    \end{subfigure}
    \begin{subfigure}[t]{0.49\textwidth}
    \includegraphics[width=1\textwidth]{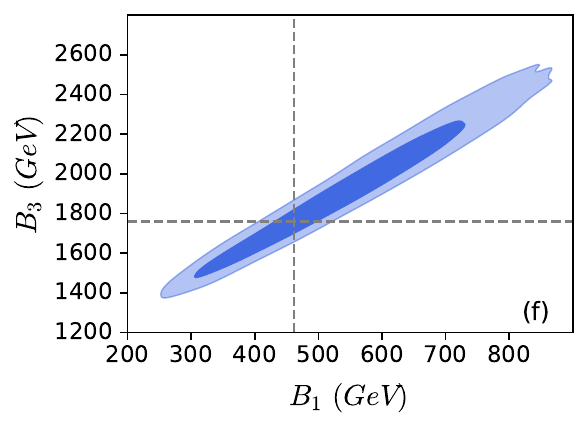}
    \label{fig:nh_B13}
    \end{subfigure}
    \caption{Marginalized posterior distribution in the (a) $\epsilon_1$-$\epsilon_2$, (b) $\epsilon_1$-$\epsilon_3$, (c) $v_1$-$v_2$, (d) $v_1$-$v_3$ (e) $B_1$-$B_2$ (f) $B_1$-$B_3$ planes. The dark blue and light blue regions represent contours at 68\% and 95\% $C.L.$
    The dashed grey lines indicate the best-fit values for both parameters.}
    \label{fig:nh1}
\end{figure}

The marginalized posterior distribution 
% our results in the form of 1$\sigma$ and 2$\sigma$ allowed regions obtained from the marginalized 2-D distributions 
for various input parameter values are presented in Fig.~\ref{fig:nh_mu_tanbeta} and Fig.~\ref{fig:nh1}.
The dark blue and the light blue regions in Fig.~\ref{fig:nh_mu_tanbeta},  represent the 68\% and 95\% confidence contours in $\mu -\tan\beta$ plane. This result is subjected to the specific choices of $A_t = -3.5$ TeV and $M_A = 3$ TeV as mentioned before. Clearly, even the $2\sigma$ allowed region for $\tan\beta$ is highly constrained. Note that, this stringent constraint is entirely due to neutrino oscillation data which proves to be far stricter than the flavor and Higgs sector observables. $\tan\beta$ affects the tree and loop contributions to the light neutrino masses in a contrasting manner as discussed in Sec.~\ref{sec:model} and is expected to be highly constrained given the small margins of uncertainty in the neutrino oscillation data.

Figure~\ref{fig:mass_split} represents how the two light neutrino mass squared differences vary with $\tan\beta$ when all other parameters are kept fixed at their best-fit values. It clearly shows that there is only one region of
$\tan\beta$ where both $\Delta m^2_{21}$ and $\Delta m^2_{31}$ can be fit simultaneously within their $2\sigma$ allowed ranges. Understandably, the allowed $\tan\beta$ here is just a number (12.11, i.e., best-fit value), and if one varies the other parameters, 
the allowed range at 95\%  $C.L.$ is obtained as shown in Figure~\ref{fig:nh_mu_tanbeta}. The allowed regions for $\tan\beta$ in the $\mu-\tan\beta$ plane at 68\% and 95\% $C.L.$ are $\sim$ [10.7-15.3] and [9.1-18.0] respectively. The choice of $\mu$ is also mostly restricted from the neutrino sector data. Similarly, the 68\% and 95\% $C.L.$ allowed ranges of $\mu$ found out to be around [1075-1460] and [1000-1510] GeV respectively for the fixed set of other parameters mentioned in Sec.\ref{sec:num}.

%%%%%%%%%%%%%%%%%%%%%%%%%%%%%%%%%%%%%%%
%%%%%%%%%%%%%%%%%%%%%%%%%%%%%%%%%%%%%%

Fig.~\ref{fig:nh1} shows the 2D marginalized posterior probability distributions in (a) $\epsilon_1$-$\epsilon_2$, (b) $\epsilon_1$-$\epsilon_3$, (c) $v_1$-$v_2$, (d) $v_1$-$v_3$, (e) $B_1$-$B_2$, (f) $B_1$-$B_3$ plane along with the allowed regions at 68\% $C.L.$ and 95\% $C.L$. The heaviest of the three neutrinos gets its mass at tree level with some additional correction from the loop diagrams. This neutrino mass is, therefore, mostly driven by the $\epsilon$ parameter. The heaviest of the three in this case has to be dominantly $\tau$-flavored. As a result, the $\epsilon_3$ and $v_3$ are expected to be the largest among the respective $\epsilon_i$ and $v_i$ parameters. On the other hand, first generation parameters $\epsilon_1$ and $v_1$ are expected to be the smallest since the heaviest neutrino has next to zero admixture of electron neutrino (see Fig.\ref{fig:nh1}(a)-(d)). The second heaviest neutrino gets most of its contribution from the $BB$-loop and it has a comparable admixture of all three neutrino flavors. As a result, we can observe nice correlations between two $B_i$'s as shown in Fig.~\ref{fig:nh1}(e) and Fig.\ref{fig:nh1}(f). These contributions are already loop-suppressed. In addition to that the value of $\tan\beta$ is also restricted from tree-level calculation by the smallness of neutrino mass scale (see Sec.\ref{sec:model}). Hence for these contributions to neutrino masses to be significant, the $B_i$ parameters have to be much larger compared to $\epsilon_i$ parameters as evident from Fig~\ref{fig:nh1}. The $\epsilon B$-loop contributions are further suppressed due to their dependence on $\epsilon$. As a result, $B_1$ is expected to be relatively larger than $B_2$ since the lightest state is dominantly electron neutrino-like. 
From the marginalized posterior distribution in $\epsilon_1$-$\epsilon_2$ and $\epsilon_1$-$\epsilon_3$ plane, it is evident that all the $\epsilon_i$ are tightly constrained. We obtain the 2$\sigma$ ranges of $\epsilon_1$, $\epsilon_2$, $\epsilon_3$ as (-1.65 to 1.08)$\times10^{-2}$, (-2.30 to 0.53)$\times10^{-2}$ and (-5.76 to -0.99)$\times10^{-2}$ GeV respectively. Similarly, the 2$\sigma$ allowed regions for $v_1$, $v_2$, $v_3$ from Fig.\ref{fig:nh1} (c) and (d) are (1.13-5.11)$\times10^{-4}$ GeV, (1.85-6.07)$\times10^{-4}$ GeV and (6.81-13.6)$\times10^{-4}$ GeV. The same regions for $B_1$, $B_2$ and $B_3$ are (250-860) GeV, (0-575) GeV and (1380-2550) GeV respectively.

In Fig.\ref{fig:nh_corner} (Appendix-\ref{appendix1}), we also present the corner plot of all the input parameters, which shows all the 1D and 2D marginalized distribution of the posterior probability and reflects the 
covariances between parameters. 
As mentioned in Sec.\ref{sec:intro}, to get the light mass eigenstates as the three neutrino masses from the $7\times7$ neutralino mass matrix, the alignment between $\epsilon_i$ and $v_i$ parameters is required (i.e., $v_i \propto \epsilon_i$). This is also reflected in the corner plot in $v_i$-$\epsilon_i$ plane (for example the first column-third row plot in Fig.~\ref{fig:nh_corner} from the top represents the correlation in $v_1$-$\epsilon_1$ plane). We have also compared our results with the previous analysis in the bRPV scenarios. In most of the previous works~\cite{Hirsch:2000ef,Gozdz:2008zz,Hundi:2011si,Hirsch:2000jt,Abada:2001zh,Diaz:2004fu}, the authors had analyzed the mSUGRA scenarios considering the then available neutrino data and these works were published before the Higgs discovery era. Among these analyses, relevant parameter space scanning was presented in Ref.~\cite{Hirsch:2000ef,Gozdz:2008zz} with contemporary neutrino data and our best-fit point along with 2$\sigma$ allowed regions have significant overlap with the previous results.
While deriving these results, all other SUSY parameters are kept fixed at values mentioned at the beginning of Sec.~\ref{sec:parameter}. Varying the SUSY parameters further can alter the results. For illustration, we have 
presented the results with different choices of $M_A$ and $A_t$ 
in Appendix.~\ref{appendix3}.   

%\tcm{Note that, in the alignment of $\epsilon_i$ parameters and sneutrino vev, which arises naturally in the framework of horizontal symmetries, the three light mass eigenstates of $L_i$ correspond to the three light neutrinos ~\cite{Banks:1995by,Allanach:2003eb}. To achieve this, the alignment between $B_i$ and $\epsilon_i$ is also required i.e., $B_i \propto \epsilon_i$~\cite{Banks:1995by,Allanach:2003eb}. In such cases, both the bilinear term and the soft breaking bilinear term can be rotated away by the field redefinition of $L_i$ and $H_u$.}

%%%%%%%%%%%%%%%%%%%%%%%%%%%%%%%%%%%%%%%%%%%%%%%%%%%%%%%%
%%%%%%%%%%%%%%%%%%%%%%%%%%%%%%%%
\subsection{Inverted Hierarchy: a comparison}  
\label{subsec:ih}
%%%%%%%%%%%%%%%%%%%%%%%%%%%%%%%%

\begin{table}[htpb!]
    \centering
	\begin{tabular}{||l|l|l||l|l||}
	\hline\hline
	 \multicolumn{3}{||c||}{Input parameters} & \multicolumn{2}{c||}{Output observables}	\\
	\hline
     Para- & Best-fit & \hspace{2mm} Mean value [95$\%$C.L.]  & Observable & Best-fit \\ 
    meter & value &  & & value \\ 
	\hline\hline
	$\epsilon_1$ & -0.0216 & -0.0199 [-0.0254, -0.0066] & $\Delta m_{21}^2$[eV$^2$] & 7.48$\times10^{-5}$  \\
	\hline
	$\epsilon_2$ & -0.0833 & -0.0803 [-0.0928, -0.0709] & $\Delta m_{31}^2$[eV$^2$] & 2.46$\times10^{-3}$ \\
	\hline
	$\epsilon_3$ & -0.0499 & -0.0479 [-0.0569, -0.0368] & $\theta_{13}$ & 8.59$^{\circ}$  \\
	\hline
	$v_1$  & 0.00086 & 0.00083 [0.00073, 0.00094] & $\theta_{12}$ & 34.23$^{\circ}$ \\
	\hline
	$v_2$ & 0.00140 & 0.00143 [0.00126, 0.00159] & $\theta_{23}$ & 49.51$^{\circ}$ \\
	\hline
	$v_3$ & 0.00110 & 0.00110 [0.00096, 0.00126] & $m_h$ [GeV] & 123.91\\
	\hline
	$B_1$ & 894.09 & 996.09  ~[764.76, 1248.98] & $BR(B \rightarrow X_s \gamma)$ & 3.16$\times10^{-4}$  \\
	\hline 
	$B_2$ & 982.76 & 959.44~ [762.32, 1184.77] & $BR(B_s \rightarrow \mu^+ \mu^-)$ & 3.21$\times10^{-9}$  \\
	\hline
	$B_3$ & 1609.81 & 1515.77 [1335.61, 1760.13] & $k_z$ & 1.0  \\
	\hline
	$\mu$ & 1437.83 & 1452.09 [1324.13, 1587.00] & $k_w$ & 1.0  \\
	\hline
	$\tan\beta$ & 8.72 & 8.25 ~~~~[6.57, 9.54] & $k_b$ & 1.001823 \\
	\hline
	&  &  & $k_{\tau}$ & 1.001823 \\
	\hline
	 &  &  &  $k_{\mu}$ & 1.001823 \\
	\hline
	 &  &  &  $k_t$ & 0.999976  \\
	\hline
	 &  &  & $k_{\gamma}$ & 1.078263 \\
	\hline\hline
	\multicolumn{5}{|c|}{ $\chi^2_{min} = $ 3.38~~~~~~~~~~~ ${\chi^2_{min}}/${DoF} = 0.845} \\
	\hline\hline
		\end{tabular} 
		\caption{Best-fit, mean values along with the 95$\% C.L.$ of all the free parameters and best-fit values for the observables in the NH scenario are shown here.  The last row represents $\chi^2_{min}$ and ${\chi^2_{min}}/${DoF} .}
	\label{tab:ih_input_output}
\end{table}

%%%%%%%%%%%%%%%%%%%%%%%%%%%%%%%%%%%%%%%%%

%%%%%%%%%%%%%%%%%%%%%%%%%%%%%%%%%%%%%%%%%
\begin{figure}[!htpb]
\begin{center}
    \includegraphics[width=0.7\textwidth]{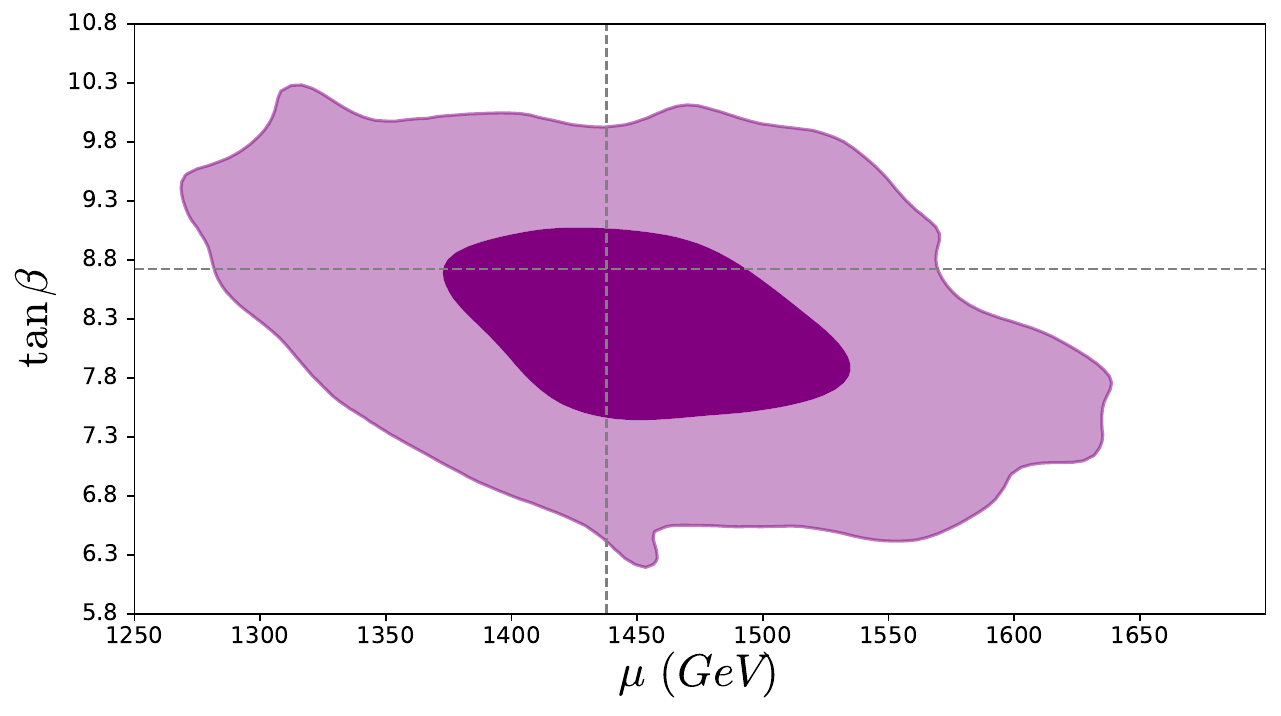}
    \caption{Marginalized posterior distribution with 68\% (dark purple) and 95\% (light purple) $C.L.$ contours in the $\mu -\tan\beta$ plane for IH scenario. The dashed grey lines indicate the best-fit values.}
    \label{fig:ih_mu_tanbeta}
\end{center}
\end{figure}
%%%%%%%%%%%%%%%%%%%%%%%%%%%%%%%%%%%%%%%%%

We perform a similar analysis assuming that the light neutrino masses obey an inverted hierarchy. The choices of input parameters and ranges are the same as the previous analysis and details are given in Sec.~3.3.
In Table \ref{tab:ih_input_output}, we present the best-fit and mean values along with the 95\% $C.L.$ allowed regions of individual input parameters. The observable values for the best-fit point are also listed in the last two columns of Table~\ref{tab:ih_input_output}. 
The best-fit point for IH corresponds to $\chi^2_{min}$ = 3.38,  
DoF = 4 and ${\chi^2_{min}}/${DoF} = 0.845. 
The contribution to the $\chi^2_{min}$ from the neutrino observables for the best-fit point is small, 0.133. The neutrino masses at the above mentioned best-fit point are $m_{\nu_1} = 4.96\times10^{-2}$ eV, $m_{\nu_2} = 5.03\times10^{-2}$ eV and $m_{\nu_3} = 1.23\times10^{-5}$ eV and their sum is $\sum_{i=1}^{3} m_{\nu_i} \approx 0.1$ eV. This evades the 2$\sigma$ upper limit on the sum of neutrino masses coming from the cosmological data i.e., $\sum_{i=1}^{3} m_{\nu_i} < 0.15 $ eV \cite{deSalas:2020pgw}.

%The marginalized posterior distribution for various input parameter values is presented in Fig. The dark blue and the light blue regions in Fig.~\ref{fig:nh_mu_tanbeta},  represent the 68\% and 95\% confidence contours in the $\mu -\tan\beta$ plane. }

%%%%%%%%%%%%%%%%%%%%%%%%%%%%%%%%%%%%%%%%%
\begin{figure}[!htpb]
    \begin{subfigure}[t]{0.49\textwidth}
    \includegraphics[width=1\textwidth]{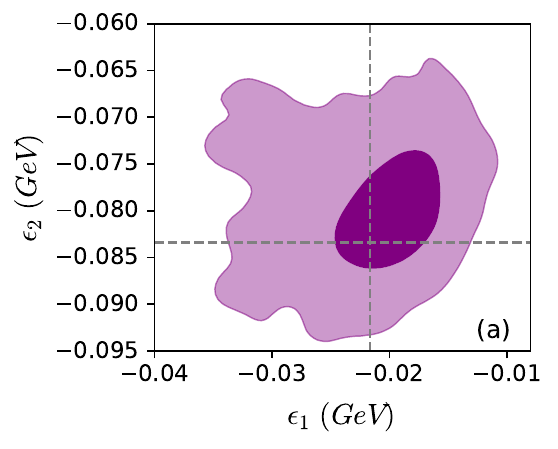}
    \label{fig:1}
    \end{subfigure}
    \begin{subfigure}[t]{0.49\textwidth}
    \includegraphics[width=1\textwidth]{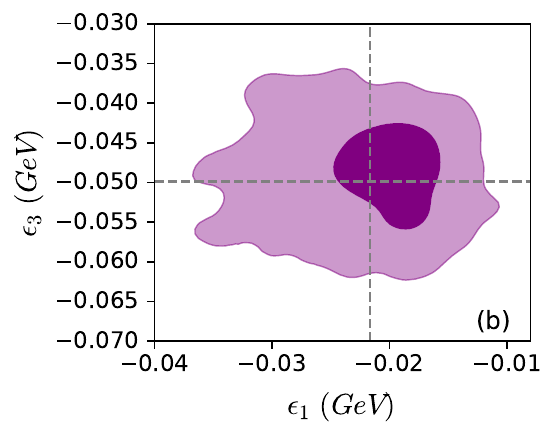}
    \label{fig:2}
    \end{subfigure}
    \\[\smallskipamount]
     \begin{subfigure}[t]{0.49\textwidth}
    \includegraphics[width=1\textwidth]{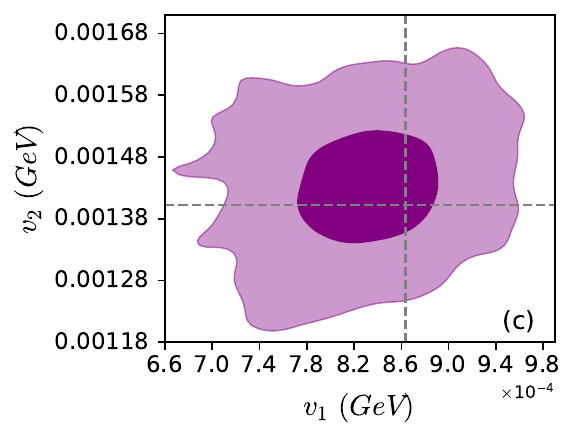}
    \label{fig:3}
    \end{subfigure}
    \begin{subfigure}[t]{0.49\textwidth}
    \includegraphics[width=1\textwidth]{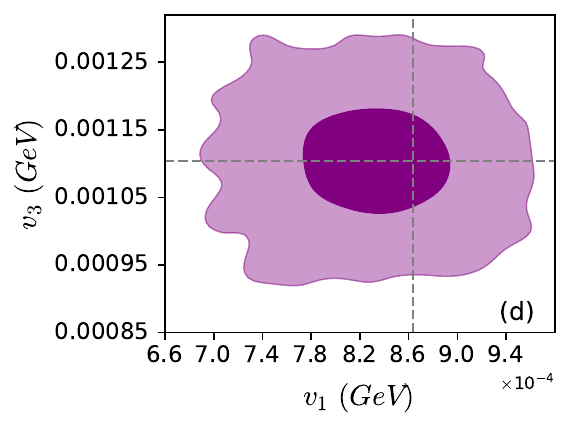}
    \label{fig:4}
    \end{subfigure}
    \begin{subfigure}[t]{0.49\textwidth}
    \includegraphics[width=1\textwidth]{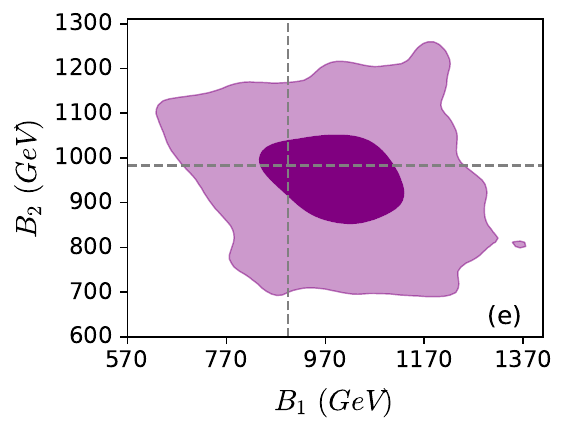}
    \label{fig:5}
    \end{subfigure}
    \hfill
    \begin{subfigure}[t]{0.49\textwidth}
    \includegraphics[width=1\textwidth]{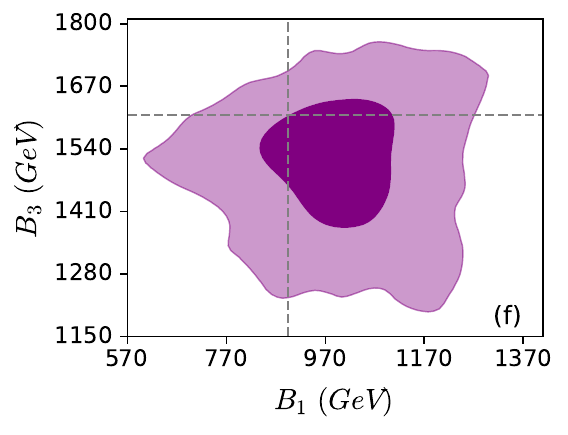}
    \label{fig:6}
    \end{subfigure}
      \caption{Marginalized posterior distribution in the (a) $\epsilon_1$-$\epsilon_2$, (b) $\epsilon_1$-$\epsilon_3$, (c) $v_1$-$v_2$, (d) $v_1$-$v_3$ (e) $B_1$-$B_2$ (f) $B_1$-$B_3$ planes. The dark and light purple regions represent contours at 68\% and 95\% $C.L.$ respectively. The dashed grey lines indicate the best-fit values for both parameters.}
    \label{fig:ih1}
\end{figure}
%%%%%%%%%%%%%%%%%%%%%%%%%%%%%%%%%%%%%%%%%

The resultant contour plot in the $\mu-\tan\beta$ plane is shown in Fig.~\ref{fig:ih_mu_tanbeta} and the dark (light) purple regions correspond to 
68\% (95\%) confidence contours. 
The 2$\sigma$ allowed regions for $\mu$ and $\tan\beta$  are [1270-1635] GeV 
and [6.2-10.2] respectively. The heaviest neutrino state in this scenario contains an almost equal admixture of all the three neutrino flavors. Tree level contribution from any single flavor cannot be very large and as a result the mixing of any single flavor neutrino with neutralino states typically would have to be smaller compared to that in a normal hierarchy scenario. That explains the comparatively smaller $\tan\beta$ and heavier $\mu$ values obtained in this case following Eq.~\ref{eq:total_mass} and Eq.~\ref{eq:XT}.

The marginalized posterior distributions of the RPV parameters are shown in Fig.~\ref{fig:ih1} with the same color convention as the previous figure. In the IH scenario, the lightest state is dominantly $\tau$-flavored with substantial admixture from $\mu$ flavor as well. The hierarchy of the other two mass eigenstates remains the same as in normal hierarchy. However, their physical masses are now larger compared to the NH case. This leads to substantial changes in the mixing pattern of the light neutrinos. Take the heaviest state as an example. Unlike in the NH scenario, this state has an almost equal admixture of all three neutrino flavors. In addition to that, the mass of this state is larger than the corresponding state in NH. This explains the slightly different ranges of the $\epsilon_i$ and the $v_i$ parameters. The hierarchy in values depends upon the relative contributions required in the admixture of the states. The ranges of $B_1$ and $B_2$ are quite different from the NH scenario (see Fig.\ref{fig:nh1}(e)-(f) and Fig.\ref{fig:ih1}(e)-(f)). Again, this has to do with the different amount of loop correction required from different generations depending on the changed hierarchy. Take the state $\nu_1$ for example. In both scenarios, the mass of this state is generated at one loop, but in the IH scenario, the physical mass of this state is larger than that in the NH scenario. This requires larger loop contributions, which explain the larger values required for $B_1$ and $B_2$.   
The 2D marginalized distributions suggest that the allowed regions are more constrained in the IH scenario compared to the NH scenario. We present the corner plot for all the input parameters i.e., all the 1D and 2D marginalized distribution of the posterior probability in Fig.\ref{fig:ih_corner} of Appendix-\ref{appendix2}.

\section{Addressing Anomalous Muon magnetic moment}
\label{sec:muon_g}
%%%%%%%%%%%%%%%%%%%%%%%%%%%%%%%%%%%%

We now proceed with our allowed parameter region to explore the possibility of explaining the existing $4.2\sigma$ excess (see Eq.\ref{eq:muon_g}) in the measurement of muon~(g~-~2)~\cite{Muong-2:2006rrc,Muong-2:2021ojo}. The excess contribution arising from both RPC and RPV processes is denoted as $\Delta a_{\mu}$. The RPC SUSY parameter space has been studied widely\tcb{\footnote{ There are only a very few analyses in the context of RPV SUSY with LQD and/or LLE couplings \cite{Altmannshofer:2020axr,Zheng:2021wnu,Chakraborty:2015bsk}, 
UDD couplings \cite{Chakraborti:2022vds}
 and bRPV scenario \cite{Hundi:2011si}.}} \cite{Baer:2021aax,Athron:2021iuf,Endo:2021zal,He:2023lgi,Chakraborti:2021bmv,Choudhury:2017acn,Choudhury:2017fuu,Choudhury:2016lku,Chakraborti:2015mra,Chakrabortty:2015ika,Chakraborti:2014gea, Banerjee:2018eaf,Banerjee:2020zvi,Chakraborti:2021dli,Frank:2021nkq,Ali:2021kxa,Chakraborti:2022vds}. We aim to explore whether the additional contributions arising in presence of RPV couplings can make a difference. As mentioned earlier, the new loop contributions that we obtain in the present scenario are through the mixing of sneutrino-higgs, neutralino-neutrino and chargino-charged lepton states. 
However, given the allowed regions of parameter space obtained in our study, it is evident that these new contributions will always be subleading because they depend on the $\epsilon_i$ and $v_i$ parameters which have to be quite small to address experimental observations.

%%%%%%%%%%%%%%%%%%%%%%%%%%%%%%%%%%%%%%%%%%%%%%%%%%%%
\begin{table}[htpb!]
   \centering
   \small	
	\begin{tabular}{||l|l|l||l||l|l||}
	\hline\hline
	\multicolumn{3}{|c||}{Input parameters} & \multicolumn{3}{c|}{Output observables} \\
	\hline\hline
	Parameters & BP-I & BP-II & Output & BP-I & BP-II \\
	\hline
	$M_1$[GeV] & 128 & 183 & $m_{\tilde{\chi}_1^0}$[GeV] & 125 & 180  \\
	\hline
	$M_2$[GeV] & 1200 & 1200 & $m_{\tilde{\chi}_1^{\pm}}$[GeV] & 1198 & 1192  \\
	\hline
	$m_{\tilde{\mu}_{L}}$[GeV] & 120 & 200 & $m_{\tilde{\mu}_1}$[GeV] & 164 & 224\\
	\hline
	 $m_{\tilde{\mu}_{R}}$[GeV] & 190 & 240  & $m_{\tilde{\mu}_2}$[GeV] & 175 & 235 \\
	\hline
	$\mu$[GeV] & 1250.02 & 1237.46  & $m_h$[GeV] & 124.62 & 124.44 \\
	\hline
	 $\tan\beta$ & 13.75 & 11.94  & $\Delta m^2_{21}$[$10^{-5}$eV$^2$] & 7.51 & 7.38 \\
	\hline
	$v_1$[GeV] & 0.000283 & 0.000321  & $\Delta m^2_{31}$[$10^{-3}$eV$^2$] & 2.56 & 2.56 \\
	\hline
	$v_2$[GeV] & 0.000390 & 0.000436  & $\theta_{13}/^{\circ}$ & 8.58 & 8.50 \\
	\hline
	$v_3$[GeV]& 0.000941 & 0.000866 & $\theta_{12}/^{\circ}$ & 34.10 & 34.38 \\
	\hline
	$\epsilon_1$[GeV] & -0.0072 & -0.0064 & $\theta_{23}/^{\circ}$ & 49.50 & 49.01   \\
	\hline 
	$\epsilon_2$[GeV] & -0.0113 & -0.0103  & $BR(B \rightarrow X_s \gamma)[\times 10^{-4}]$ & 3.13 & 3.14  \\
	\hline		
	$\epsilon_3$[GeV] & -0.0446 & -0.0280 & $BR(B_s \rightarrow \mu^+ \mu^-)[\times 10^{-9}]$ & 3.22 & 3.21   \\
	\hline
	$B_1$[GeV]  & 422.20 & 467.16 & \textbf{$\Delta a_{\mu}$} [$\times  10^{-10}$] & \textbf{25.41}  & \textbf{13.52} \\
	\hline
	$B_2$[GeV] & 134.30 & 149.44 &  &  &  \\
	\hline
	$B_3$[GeV] & 1931.88 & 1989.45 &  &  &  \\
	\hline\hline
	\end{tabular}
	\caption{Details of benchmark points (BP) satisfying the muon~(g~-~2) data along with other observables.}
\label{tab:muon_g-2}
\end{table}
%%%%%%%%%%%%%%%%%%%%%%%%%%%%%%%%%%%%%%%%%%%%%%%%%%%%%%%%%%%%%%%%%%%%%%%%

One would have ideally added $\Delta a_{\mu}$ as another observable in the analysis itself to get a complete picture, but obtaining the required excess in $\Delta a_{\mu}$ also requires us to vary bino, wino as well as the slepton soft mass parameters. That would have made our analysis more complex and time-consuming. Hence we have kept our $\Delta a_{\mu}$ analysis separate and present our results in terms of some chosen benchmark points. In Table~\ref{tab:muon_g-2}, we have enlisted all the input parameters and relevant observables along with the obtained value of $\Delta a_{\mu}$.
We have tried to find benchmark points that can explain $\Delta a_{\mu}$ within the 2$\sigma$ allowed range of the global average mentioned in Eq.~\ref{eq:muon_g}. We have considered the NH scenario here for this analysis. One can obtain similar results for the IH scenario as well. 
For the benchmark points (BP-I and BP-II), we have fixed the following parameters as: $A_t$ = -3.5 TeV, $M_A$ = 3 TeV, $M_3$ = 3 TeV, 
$m_{\tilde{q}}$ = 3 TeV (all 3 generations) and $m_{\tilde{e_{L/R}}}$ = $m_{\tilde{\tau}_{L/R}}$ = 2 TeV to evade the current LHC bounds on 
the sparticle masses as mentioned in section \ref{sec:parameter}. 
The relevant smuon mass $m_{\tilde{\mu}}$ is mentioned in the 
Table~\ref{tab:muon_g-2} for the two benchmark points. BP-I (BP-II) satisfies the observed value of $\Delta a_{\mu}$ within 1$\sigma$ (2$\sigma$) limit. This difference happens due to the larger smuon mass in BP2. Note that, the new physics contributions obtained for these spectra are very similar to R-parity conserving scenario. Except for the LSP, the decay of all other SUSY particles is dominantly R-parity conserving because of the smallness of the RPV parameters. Hence the applicable collider limits for these sparticles happen to be the same as in R-parity conserving MSSM\footnote{It may be noted that our benchmark points satisfy both the muon~(g~-~2) anomaly and LHC constraints coming from slepton pair production and these results are similar to the RPC scenario considered in Ref.~\cite{Endo:2021zal} (see Fig.2B of Ref.\cite{Endo:2021zal}).} except for the LSP. The LSP decays through RPV couplings and for its mass existing RPV limits have been taken into account \cite{ATLAS:2020uer}.

%%%%%%%%%%%%%%%%%%%%%%%%%%%%%%%%%%%%%%%%%%%%%%%%%%
%Although for massless $\tilde{\chi}_1^0$, the LHC data has already excluded $m_{\tilde{\mu}_{L/R}}$ upto 600 GeV (see Fig.8(b) of Ref\cite{ATLAS:2019lff}.But the LHC data has not still able to explore the regions where the mass difference of slepton and neutralino is small. Our Benchmark points lie in this region. For example at $m_{\tilde{\chi}_1^0}$ = 125 GeV the slepton mass in the range 190-700 (600) GeV is excluded from the direct search production of first two generations sleptons (only 2nd generation i.e., smuon ($\tilde{\mu}$). \par

%\tcb {In RPC SUSY scenario universal slepton mass, the authors in Ref.\cite{Endo:2021zal} have shown that $m_{\tilde{\mu}}$ $\lesssim$ 270 GeV can explain both (g-2) anomaly and the LHC constraints coming from slepton pair production searches (See Fig. 2B of Ref.~\cite{Endo:2021zal})}

\section{Conclusion}
\label{sec:conclusion}

The existence of non-zero masses and non-trivial mixing among the light neutrino states have been established beyond any doubt by the neutrino oscillation experiments. The standard model cannot address this phenomenon within its framework and that makes this observation one of the most robust indications for new physics beyond the standard model. Any completely new physics model, therefore, should be able to address this issue. Theoretically, Supersymmetry remains one of the most well-motivated new physics scenarios till date and hence various supersymmetric scenarios have been extensively studied both by theoretical and experimental collaborations. Supersymmetric extension of the standard model with conserved R-parity requires an additional mechanism to explain the neutrino oscillation data while with bilinear R-parity violating scenario, one can easily address this long-standing issue without extending the model any further. In the absence of any robust indication of new physics at the LHC, the need of the hour is to study the existing models under the light of the plethora of experimental data at our disposal and try to constrain the new physics parameter regions as much as possible. Neutrino oscillation data is more precise and capable of constraining the new physics parameter space compared to direct search data, as shown in this paper. In addition to that, taking into account the precision Higgs data and flavor data, we have computed the most restricted bRPV SUSY  parameter space  that can be obtained
(with $\chi^2_{min}$ / DoF $\sim 1$ for both the neutrino hierarchical  scenarios) using MCMC analysis.% with existing experimental data. 
~Moreover, our analysis also considers the existing direct search limits.

Our results show that the $\epsilon_i$ and the $v_i$ parameters in particular have to be quite small which means that except for the LSP, the decays of the neutralinos and charginos are expected to be mostly similar to the R-parity conserving scenario. The $B_i$ parameters can be comparatively much larger since they only contribute to the neutrino masses and mixing angles at the loop level. Since these parameters are responsible for slepton-charged Higgs and sneutrino-neutral Higgs mixing, the phenomenology of these particles is expected to differ accordingly from the R-parity conserving scenario. Thus, the results derived here will be extremely helpful in making any future studies more focused and results more predictive. Understandably, a slight change in one of the experimental data points can change the best-fit point quoted in this paper, but unless the experimental results change too much, the $2\sigma$ allowed regions are expected to remain similar. We further proceed to address the existing anomalous muon~(g~-~2) result with the constrained parameter space obtained in this work. We find out that unless we are in the compressed region where the slepton and gaugino masses are lying close to each other, it is very difficult to abide by the collider bounds and still explain the muon~(g~-~2) excess. In this regard, our results are quite similar to what is expected in the R-parity conserving scenario since the smallness of the RPV couplings ensures that there are no new significant contributions arising because of R-parity violation.

\newpage

\noindent \textbf{Acknowledgments}

\noindent Authors would like to acknowledge S K Patra for fruitful discussions regarding the analysis.

\bibliography{reference}
\newpage

\appendix
\section{Corner plot for the NH scenario}
\label{appendix1}

%\vspace{-4cm}
\begin{figure}[!htb]
\centering
  \includegraphics[width=1.1\textwidth]{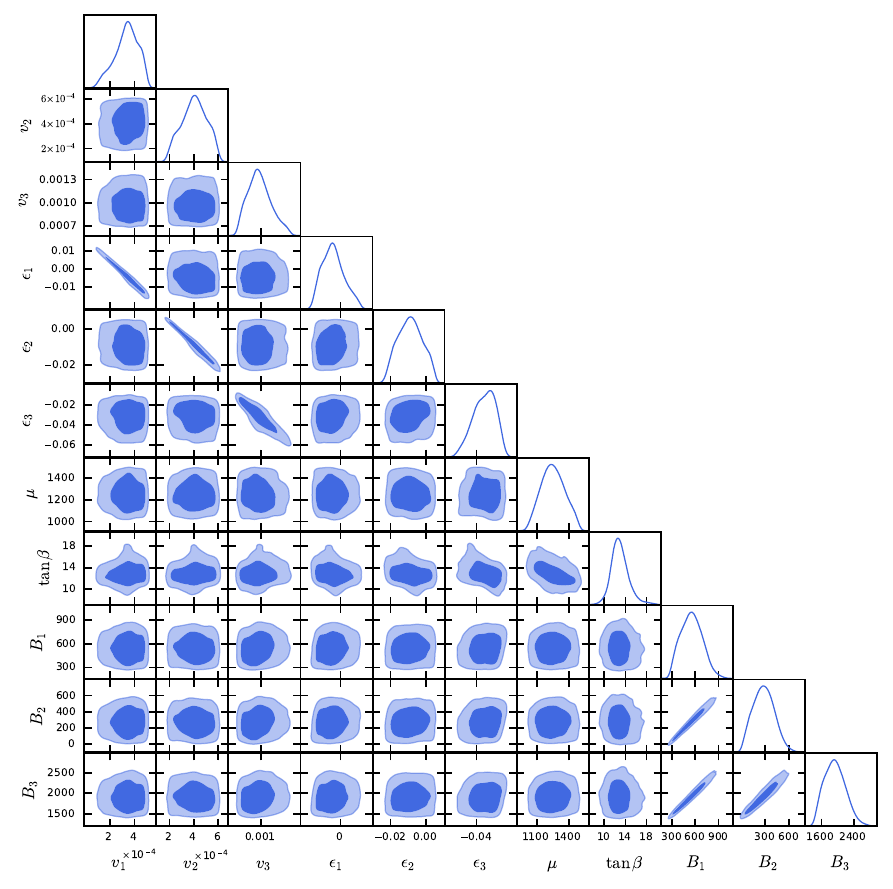}
  \caption{Corner plot for the input parameters in NH scenario. The diagonal histograms are 1D posterior probability distributions and the other contour plot show the covariances between parameters. The darker blue color contour shows 1$\sigma$ region and the light blue contour indicates 2$\sigma$ region.}
  \label{fig:nh_corner}
 \end{figure}

\newpage
 
\section{Corner plot for  the IH scenario}
 \label{appendix2}

\begin{figure}[!htb]

\centering
  \includegraphics[width=1.1\textwidth]{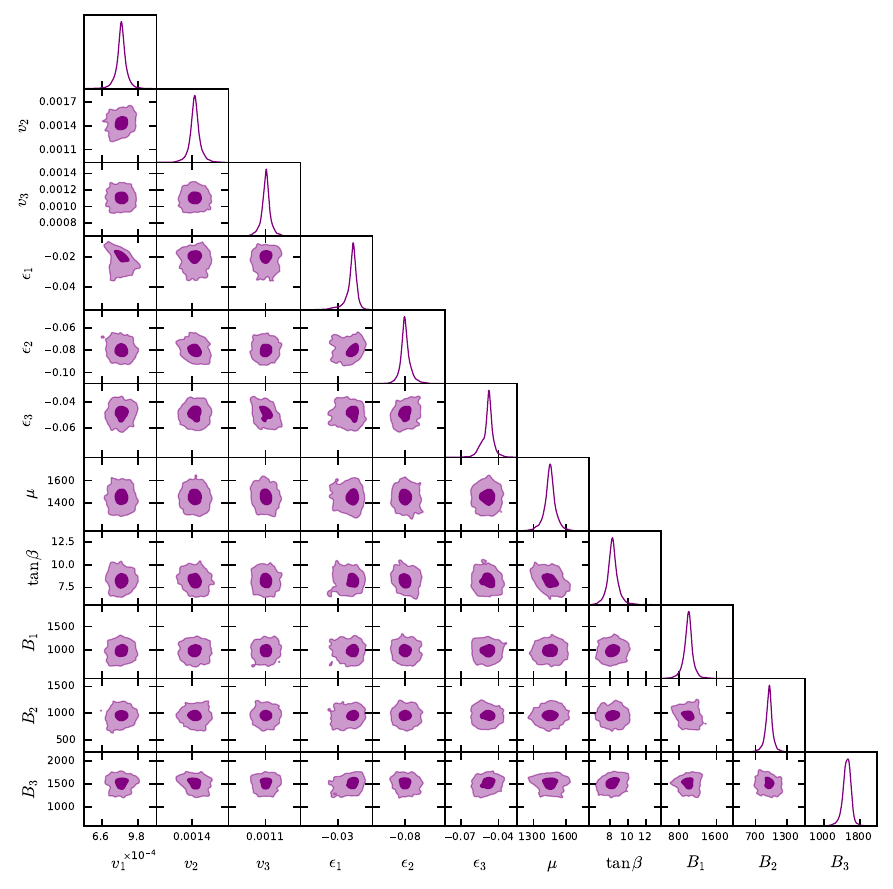}
  \caption{Corner plot for the input parameters in IH scenario. The diagonal histograms are 1D postrior probability distributions and the other contour plot show the covariances between parameters. The darker purple color contour shows 1$\sigma$ region and the light purple contour indicates 2$\sigma$ region.}
  \label{fig:ih_corner}
 \end{figure} 
 
\newpage

\section{Results corresponding to different choices of $M_A$ and $A_t$}
\label{appendix3}

The results presented in Sec.\ref{sec:result} correspond to 
$M_A$= 3 TeV and $A_t$ = -3.5 TeV. For the Normal Hierarchy scenario, 
we estimate how sensitive our results are on these choices by doubling the 
$M_A$ and $A_t$ values.  The comparisons of the best-fit, mean values along with 95\% $C.L.$ of input parameters and $\chi^2_{min}$ are shown in the Table.~\ref{tab:nh_compare}. 
It is observed that for most of the parameters, the best fit/mean values and the typical ranges are quite close in terms of order of magnitude. 
 For some of the parameters, the best-fit point has slightly shifted, e.g.,  
  $B_3$. 
 This indicates that for larger $M_A$ and $A_t$ one requires larger loop corrections for the light neutrino masses. 
It should be noted that 2$\sigma$ allowed regions with doubled $M_A$ and $A_t$ are extended a little bit more for some of the parameters due to larger $\chi^2_{min}$, which has increased to 5.01 from 3.46 and subsequently  ${\chi^2_{min}}/${DoF} has raised to 1.25 from 0.865. We observe that this is mostly due to 
the poor fitting of the flavor physics observables.

\begin{table}[!htb]
    \centering
	\begin{tabular}{||l|l|l||l|l||}
	\hline\hline
	 Para- & \multicolumn{2}{c||}{$M_A$ = 3 TeV \& $A_t$ = -3.5 TeV} & \multicolumn{2}{c||}{$M_A$ = 6 TeV \& $A_t$ = -7.0 TeV}\\
	 meter & \multicolumn{2}{c||}{Normal Hierarchy scenario} & \multicolumn{2}{c||}{Normal Hierarchy scenario} \\
	\hline
     & Best-fit & \hspace{2mm} Mean value [95$\%$C.L.] & Best-fit & Mean value [95$\%$C.L.] \\ 
	\hline\hline
	$\epsilon_1$ & -0.013 & -0.0045[-0.0183, 0.0054] & -0.018 & -0.0077[-0.025, 0.022]  \\
	\hline
	$\epsilon_2$ & -0.0160 & -0.0089[-0.0218, 0.0034] & -0.021 & -0.013~~[-0.036, -0.16] \\
	\hline
	$\epsilon_3$ & -0.0279 & -0.0311[-0.0487, -0.0068] & 0.0079 & -0.034 [-0.075, 0.0072] \\ 
	\hline
	$v_1$  & 0.00038 & 0.00034[0.00019, 0.00055] & 0.00057 & 0.00037[-0.0001, 0.00063]  \\
	\hline
	$v_2$ & 0.00052 & 0.00040[0.00022, 0.00060] & 0.00064 & 0.00047[0.00005, 0.00082] \\
	\hline
	$v_3$ & 0.00091 & 0.00097[0.00061, 0.00122] & 0.00026 & 0.00090[0.00032, 0.0015]\\
	\hline
	$B_1$ & 461.78 & 548.38~ [264.50, 791.66] & 787.77 & 857.83 ~[447.29, 1044.04] \\
	\hline 
	$B_2$ & 198.62 & 276.03 ~[5.01, 515.16] & 52.37 & 136.13 ~~[-266.61, 265.82]  \\
	\hline
	$B_3$ & 1760.66 & 1917.76[1359.46, 2355.98] & 4177.19 & 4093.42[3354.54, 4491.93] \\
	\hline
	$\mu$ & 1293.80 & 1249.33[1028.70, 1449.71] & 1446.30 & 1323.65 [960.81, 1603.67] \\
	\hline
	$\tan\beta$ & 12.11 & 12.81 ~~~[8.76, 15.67] & 9.97 & 12.44 ~~~[7.56, 15.99] \\
	\hline\hline
	& \multicolumn{2}{c||}{ $\chi^2_{min} = $ 3.46,~${\chi^2_{min}}/${DoF} = 0.865}& \multicolumn{2}{c||}{ $\chi^2_{min} = $ 5.01,~${\chi^2_{min}}/${DoF} = 1.25} \\
	\hline\hline
		\end{tabular} 
\caption{The comparison of best-fit and mean values along with 2$\sigma$ ranges corresponding to ($M_A$, $A_t$) = (3, -3.5)~TeV and (6, -7.0) TeV for the Normal Hierarchy scenario. 
	%of the different parameters are shown here. The choice with $M_A$ = 3 TeV \& $A_t$ = -3.5 TeV corresponds to the results presented for the Normal Hierarchy scenario in our manuscript (see Table 4) and the other one corresponds to the results that we have obtained from the analysis where values of $M_A$ and $A_t$ are doubled. 
	The last row represents the $\chi^2_{min}$ and $\chi^2_{min}/DoF$ for each analysis.} 
	\label{tab:nh_compare}
\end{table}

\end{document}